\documentclass[twocolumn,showpacs]{revtex4-1}

\usepackage{color}
\usepackage{graphicx}
\usepackage{epstopdf}
\usepackage{dcolumn}
\usepackage{amsmath}
\usepackage[caption=false]{subfig}

\begin{document}
\bibliographystyle{revtex}


\title[Short Title]{Symmetry energy in neutron star matter}

\author{C.O. Dorso}
\affiliation{Instituto de F\'\i sica de Buenos Aires, Pabell\'on
I, Ciudad Universitaria, 1428 Buenos Aires, Argentina. }

\author{G.A. Frank}
\affiliation{Unidad de Investigaci\'on y Desarrollo de las
Ingenier\'\i as, Universidad Tecnol\'ogica Nacional, Facultad Regional Buenos
Aires, Av. Medrano 951, 1179 Buenos Aires, Argentina. }

\author{J.A. L\'opez}
\affiliation{University of Texas at El Paso, El Paso, Texas 79968, U.S.A.}

\date{\today}

\begin{abstract}

We investigate the structure attained by neutron star matter with proton 
to neutron ratios ranging from $x=0.1$ to 0.5, densities in the range of 0.02 
fm${}^{-3}<\rho<$ 0.085 fm${}^{-3}$, and temperatures  $T<4$~MeV. In particular 
we study the pasta shapes and the phase changes previously observed in nuclear 
matter, as well as the behavior of the symmetry energy. We corroborate the 
existence of homogeneous to non-homogeneous phase transitions, and the different 
values of the symmetry energy at different densities and temperatures.

\end{abstract}

\maketitle

\section{Introduction}\label{intro}

Non-pulsar neutron stars are expected to cool down mainly through
neutrino emission. Such emission depends initially on the convective
motion of the neutron star matter, which is dictated by its equation
of state~\cite{Roberts}, its dependence on isospin content, and by
the structure of the neutron star crust which controls the opacity
of the medium to the neutrino flow \cite{dorso2017}. This 
underlines the importance of studying both the dependence of the equation of 
state (EOS) of neutron star matter on isospin content (i.e. the symmetry energy,
$E_{sym}$), and the structure of the neutron star crust at saturation and
subsaturation densities, low temperatures and with varying isospin
content.\\

The connection between the structure of a nuclear medium and the
value of its symmetry energy was established experimentally in
low-energy nuclear reactions at the Texas A$\&$M
Cyclotron~\cite{kowa,wada,natowitz}. These investigations showed
that $E_{sym}$ is affected by the formation of clusters; such
dependence on clustering was later corroborated by a calculation of
$E_{sym}$ in clustered media~\cite{lopez2017}, while  non
clustering mean-field theories failed to yield the asymptotic limit
of $E_{sym}$ at small densities~\cite{li}.\\

In a previous study, the structure and symmetry energy of nuclear matter (NM), 
 a system closely related to neutron star matter (NSM), was 
studied~\cite{dorso2018_arxiv} (NM is a conglomerate of protons 
and neutrons in which proton-proton Coulomb interaction is disregarded. NSM is 
NM with proton-proton Coulomb term activated, embedded in an electron cloud 
(the system as a whole is neutral). This electron cloud renders the Coulomb 
interaction short ranged, and then, the energy and entropy of NSM is additive). 
Such study focused on the structure of neutron-rich nuclear matter and, in 
particular, on the behavior of the symmetry energy on those structures. It was 
found, first, that the so-called ``pasta'' structures exist in neutron-rich 
nuclear matter and that they exhibit the crystal-to-solid-to-liquid phase 
transitions found in Ref.~\cite{dorso2014} for isospin-symmetric NM. And, 
second, that the symmetry energy depends on the morphology of the pasta and on
its phase transitions.\\

This motivates the questions of the present study: how does the topology of the 
NSM evolves as temperature is lowered in terms of its isospin asymmetry? Do NSM 
exhibit phase transitions similar to those observed in NM? And, how does the 
symmetry energy depend on these structures and phase changes? The importance of 
these questions lies on the potentially large impact these effects can have on 
neutrino cooling and in the synthesis of heavy nuclei in neutron 
star mergers~\cite{dorso2018}.\\

These questions can be addressed from a theoretical point of view using 
a model capable of forming the pasta structures, undergoing phase transitions, 
while allowing the evaluation of $E_{sym}$ in these cold and inhomogeneous 
systems.  Although several models exist, not all of them allow the simultaneous 
calculation of the structures, phase transitions and of the symmetry energy 
simultaneously.\\

Nuclear pastas are spatial arrangement of protons and neutrons,
theorized to exist in neutron star crusts~\cite{ravenhall}. To
obtain these structures, nuclei in the NSM should reach a free energy
minima. In the original studies of the 1980s the energy minima were
determined using static methods such as the liquid drop
model~\cite{ravenhall,Hashimoto}, mean field theories~\cite{Page}
and Thomas-Fermi models~\cite{Koonin}. These methods, however,
usually work only at zero temperature, locate only the
``traditional'' global minima and tend to miss the
``non-traditional'' local minima of energy barriers; to gain access
to the complete set of structures (traditional and non-traditional)
temperature-dependent dynamical models are needed.\\

Dynamical studies of the pastas have been performed with quantum
molecular dynamics~\cite{Maruyama,Kido,Wata2002} and classical
potential models~\cite{Horo2004,dor12, dor12A}.  But to properly
obtain the pasta structures, including the non-traditional phases,
it is best to use models that can identify local minima by cooling,
such as the classical molecular dynamics model (CMD)~\cite{dorso2014}.\\

The CMD has been used, for instance, to calculate $E_{sym}$ in 
clustered media~\cite{lopez2017}, and to study the pasta structures 
and phase changes in NM, as well as the symmetry energy in the 
different pasta structures and in all of its phases~\cite{dorso2014}.\\

In this work we extend the study of nuclear matter of
Ref.~\cite{dorso2018_arxiv} to the realm of neutron star matter. We 
use CMD to study, first, the existence of pasta-like structures and their 
possible phase changes at saturation and sub-saturation densities, low 
temperatures, and proton fractions in the range of 10$\%$ to 
50$\%$. Second, we compute the behavior of $E_{sym}$ in the different 
structures found.\\

\section{Classical molecular dynamics}\label{NSP}

This work uses classical molecular dynamics, a well-established computational 
technique that has been used to determine the pasta structures that form in NM 
and NSM~\cite{dor12, dor12A, dorso2014, lopez2014, dorso2017}.  CMD 
represents nucleons as classical particles interacting through pair potentials 
and calculates their dynamics by solving their equations of motion 
numerically.  CMD has several advantages that have been presented
elsewhere~\cite{dor12,dor12A,dorso2014}, and its validity on nuclear systems at 
the range of temperatures and densities achieved in intermediate-energy nuclear 
reactions has been discussed before~\cite{lopez2014}.\\

In this section we introduce the nucleon-nucleon potentials used in CMD 
along with the tools used in the study of the resulting structures, namely, 
the Minkowski functionals and the symmetry energy. A more thorough 
description of these tools can be found in~\cite{dorso2018_arxiv}.

\subsection{Potentials}\label{cmd}

Neutron star matter is composed of protons, neutrons and electrons, CMD uses 
$pp$, $nn$ and $np$ potentials, as well as an screening potential to mimic the effect 
of the electron gas; these potentials are now described in turn.\\

\subsubsection{nucleon-nucleon potentials}\label{cmd_star-1}

\begin{figure*}[!htbp]
\centering
\subfloat[$V_{nn}$\label{fig:nn}]{
\includegraphics[width=\columnwidth]
{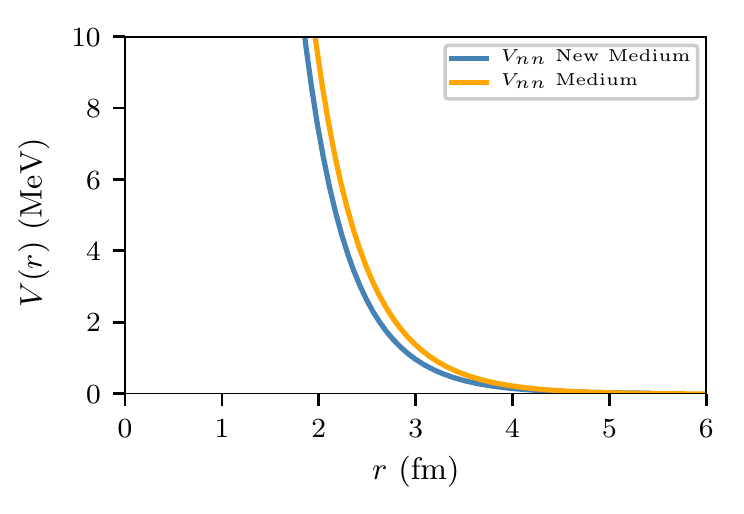}
}
\subfloat[$V_{np}$\label{fig:np}]{
\includegraphics[width=\columnwidth]
{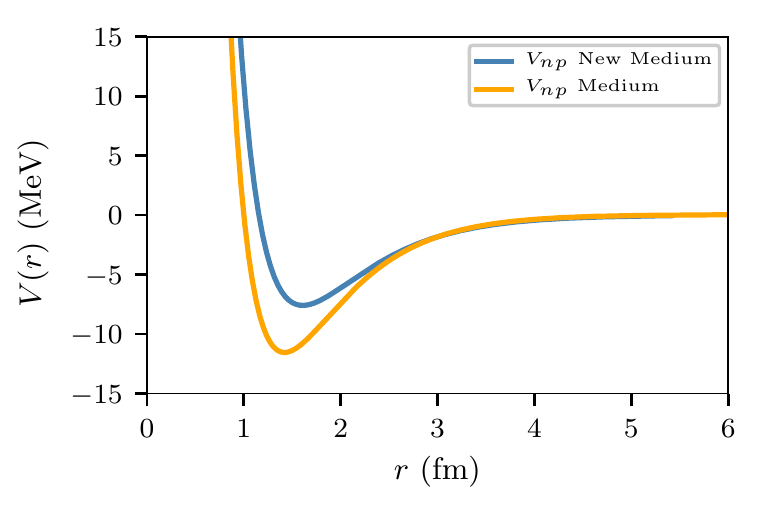}
}
\caption{\label{fig:pandha} (color online) Potential profiles for
interacting nucleons (in MeV). The orange curves correspond to the Pandharipande
Medium model. The blue curves correspond to the improved
New Medium model (see text for details).   }
\end{figure*}

In the CMD model nucleons interact through the Pandharipande (Medium) 
potentials. These potentials attain a binding energy $E(\rho_0)=-16$ 
MeV/nucleon and a compressibility of about $250\,$MeV. The corresponding 
mathematical expressions are

\begin{equation}
\begin{array}{rcl}
        V_{np}(r) & = &
\displaystyle\frac{V_{r}}{r}e^{-\mu_{r}r}-\displaystyle\frac{V_{r}}{r_c}e^{-\mu_
{ r } r_ { c } } -\displaystyle\frac{V_ { a }}{r}
e^{-\mu_{a}r}+\displaystyle\frac{V_{a}}{r_{c}}e^{-\mu_{a}r_{c}}\\
       & & \\
       V_{nn}(r) & = &
\displaystyle\frac{V_{0}}{r}e^{-\mu_{0}r}-\displaystyle\frac{V_{0}}{r_{c}}e^{
-\mu
_{0}r_{c}}
       \end{array}
\end{equation}

\noindent where $r_c$ is the cutoff radius after which the
potentials are set to zero. Although the parameters $\mu_r$,
$\mu_a$, $\mu_0$ and $V_r$, $V_a$, $V_0$ were first set by
Pandharipande for cold nuclear matter~\cite{pandha}, a recent 
improvement~\cite{dorso2018}, here named \textit{New Medium}, 
reproduces the cold nuclear matter binding energies more accurately 
and, thus, is used in this work. The corresponding values are 
summarized in Table~\ref{table_parameter}. Figs.~\ref{fig:nn} and \ref{fig:np}
contrasts these potentials with those of Pandharipande Medium potentials.


\begin{table}
{\begin{tabular}{l @{\hspace{6mm}}@{\hspace{6mm}} r
@{\hspace{11mm}} r @{\hspace{6mm}} l}
\toprule
      Parameter &  \multicolumn{1}{l}{Pandharipande} & $\ $New Medium  & 
\multicolumn{1}{l}{Units}
\\
\colrule
$V_r$ &  3088.118 & 3097.0  & MeV \\
$V_a$ &  2666.647 & 2696.0  & MeV\\
$V_0$ &  373. 118 & 379.5  & MeV\\
$\mu_r$ & 1.7468 &  1.648 & fm$^{-1}$ \\
$\mu_a$ & 1.6000 &  1.528 & fm$^{-1}$ \\
$\mu_0$ & 1.5000 &  1.628 & fm$^{-1}$ \\
$r_c$   & 5.4    &  5.4/20 & fm \\
\botrule
\end{tabular}
}
\caption{Parameter set for the CMD computations. The values used in 
this work correspond to the New Medium Model. }
\label{table_parameter}
\end{table}

\subsubsection{The Coulomb potential}\label{cmd_star-2}

As the neutron star matter contains protons and neutrons embedded in an electron 
gas, it is necessary to include the Coulomb effect of the gas. The main effect of the 
electron cloud is to introduce an screening effect on the Coulomb potential of the protons. 
The screened Coulomb potential may be expressed through the Thomas-Fermi  
expression~\cite{dor12,Horo2004,Maruyama}

\begin{equation}
V_{tf}=\displaystyle\frac{q^2}{r}e^{-r/\lambda},\label{eq:coulomb}
\end{equation}

\noindent where the screening length is $\lambda=20\,$fm, and the 
cutoff distance for $V_{tf}$ is 20 fm.  The values were selected to be long 
enough to reproduce the density fluctuations in the cell size used (see Section~\ref{MDSNSM})~\cite{dor14} .

\begin{figure}
\begin{center}
   \includegraphics[width=\columnwidth]{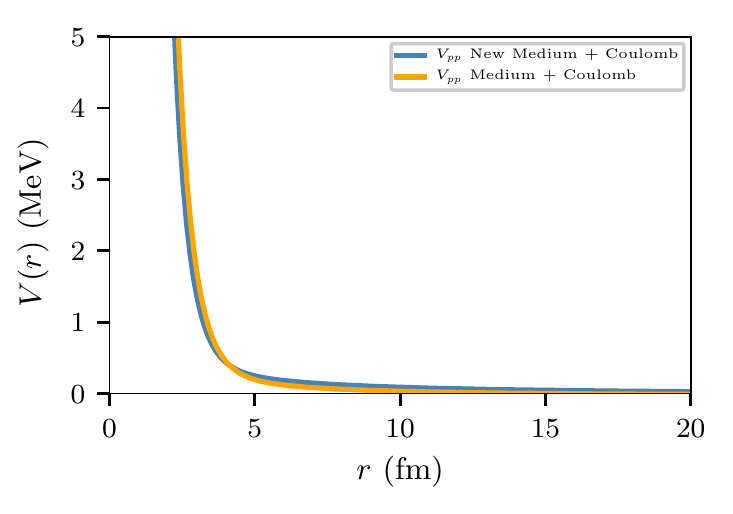}
\caption{(color online) Potential profiles for
interacting protons (in MeV). The orange curve corresponds to the
Pandharipande Medium model embedded in the Thomas-Fermi potential 
(see Eq.~(\ref{eq:coulomb})). The blue curve corresponds to 
the improved New Medium model embedded in the Thomas-Fermi potential 
(see Eq.~(\ref{eq:coulomb})).}\label{fig:coulomb}
\end{center}
\end{figure}

Figs.~\ref{fig:pandha} shows the interaction potentials between nucleons,
without the existence of the surrounding electron gas, while Fig.~\ref{fig:coulomb} 
shows proton-proton complete potential (including the Coulomb screening).

It might be appropriate at this point to emphasize that the 
exponential cut-off renders the Coulomb effective interaction, short ranged. In 
this way the energy and the entropy is additive (\textit{i.e.} energy scales 
with the number of particles).

\subsection{Tools}\label{tools}

To characterize the pasta here we use the Minkowski functionals~\cite{michielsen}, which 
allow us to measure the size, shape and connectivity of spatial structures formed by the 
nucleons. In this section we also present the procedure used for the evaluation of the symmetry 
energy from CMD data~\cite{lopez2014,lopez2017}. 

\subsubsection{Minkowski functionals}\label{mink}

The Minkowski functionals~\cite{michielsen} are shape descriptors that have been used extensively 
in the pasta literature (see e.g.~\cite{Watanabe2003,dor12A,Schneider,Caplan}). For three-dimensional 
bodies, these functionals are the volume, surface area, Euler characteristic ($\chi$), and
integral mean curvature (B). The Euler characteristic can be interpreted as
\begin{equation}
\chi = \mathrm{isolated\ regions} + \mathrm{cavities} -
\mathrm{tunnels}\label{eq:chi}
\end{equation}

\noindent while $B$ is a measure of the curvature of the surface of a given
structure. In Ref.~\cite {dor12A} it was found that the pasta
structures can be classified according to Table~\ref{tab1}, i.e. a characterizing connection exists 
between the different structures with the curvature (B) and the Euler characteristic; the signs of 
these functionals are also correlated with the pasta phases~\cite{Schuetrumpf2013}.  In our case, 
to calculate the Minkowski functionals on pasta structures produced by CMD, the structures must be 
``voxelized'', i.e., represented by ``voxels'', or regular units of volume; the appendix 
of~\cite{dorso2018_arxiv} presents the procedure used.

\begin{table}
{\begin{tabular}{c @{\hspace{4mm}}|@{\hspace{4mm}} c
@{\hspace{4mm}}|@{\hspace{4mm}} c @{\hspace{4mm}}|@{\hspace{4mm}} c}
\toprule
           &   B $<$ 0  & B $\sim$ 0  &   B $>$ 0 \\
\colrule
$\chi > 0$ & Anti-Gnocchi &  & Gnocchi\\
$\chi \sim 0$ & Anti-Spaghetti  & Lasagna & Spaghetti\\
$\chi < 0$ & Anti-Jungle Gym &  & Jungle Gym \\
\botrule
\end{tabular}
}
\caption{Integral mean curvature and Euler characteristic values for
pasta shapes. The ``anti'' prefix means the inverted situation between
occupied and empty regions. The ``jungle gym'' stands for a 3D rectangular
trellis.}
\label{tab1}
\end{table}

\subsubsection{Symmetry energy}\label{esymm}

The evaluation of the symmetry energy follows the procedure
introduced in Refs.~\cite{lopez2014,lopez2017}. The symmetry energy
is defined as

\begin{equation}
E_{sym}(\rho,T)=\displaystyle\frac{1}{2!}\,\displaystyle
\frac{\partial^2E(\rho,T,\alpha)}{\partial
\alpha^2}\bigg|_{\alpha=0}\label{esym_definition}
\end{equation}

\noindent with $\alpha=(N-Z)/(N+Z)=1-2x$. Using the CMD results of the internal
energy $E(\rho,T,x)$ it is possible to construct a continuous function by
fitting (\textit{i.e.} least squares estimation) the values of
$E(T,\rho,\alpha)$ for each $T$ and $\alpha$ with an expression of the type

\begin{equation}
E(T,\rho,\alpha)=
\displaystyle\sum_{i=0}^3E_i(T,\alpha)\,\rho^i\label{fitting_1}
\end{equation}

The $\alpha$ dependence of the coefficients $E_i(T,\alpha)$ can be extracted
from the CMD data calculated at various values of $\alpha$, and assuming an
$\alpha$ dependence of the type

\begin{equation}
E_i(T,\alpha)=E_{i0}(T)+E_{i2}(T)\,\alpha^2+E_{i4}(T)\,\alpha^4\label{fitting_2}
\end{equation}

\noindent with odd terms in $\alpha$ not included to respect the isospin
symmetry of the strong force. The symmetry energy is then given by

\begin{eqnarray}
E_{sym}(T,\rho)&=&E_{02}(T)+ E_{12}(T)\rho+ \nonumber
\\ &+&E_{22}(T)\rho^2+E_{32}(T)\rho^3
\end{eqnarray}

\noindent with the coefficients $E_{ij}(T)$ obtained from the fit of the CMD
data.

\subsection{Molecular dynamics simulation of neutron star matter}\label{MDSNSM}

In order to mimic (asymmetric) neutron star matter, the LAMMPS CMD code~\cite{lammps} was
fitted with the potentials mentioned in Sections~\ref{cmd_star-1} and 
\ref{cmd_star-2}. Graphics Processing Units (GPU) were used to carry out the 
computations. We tracked the evolution of systems with $A=$ 4000 nucleons 
situated in a cubic cell under periodic boundary conditions. The simulated 
isospin content was $x = z/A= 0.1$, $0.2$, $0.3$, $0.4$, and 
$0.5$. The densities were varied between $0.02\,$fm$^{-3}$ to 
$0.085\,$fm$^{-3}$. The temperature was controlled with a Nos\'e-Hoover 
thermostat slowly varying from $T$ = 4 MeV down to 0.2 MeV ($\Delta T < 0.1\%$). 
After placing the nucleons at random, but with a minimum inter-particle distance 
of $0.01\,$fm, the nucleons were endowed with velocities according to a 
Maxwell-Boltzmann distribution to correspond to a desired temperature, and the 
equations of motion were solved to mimic the evolution of the system. The 
nucleon positions, momenta, energy per nucleon, pressure, temperature, and 
density, were stored at fixed time-steps.\\

\section{Results for NSM pasta}\label{results_1}

With the tools presented in the previous section we now analyze the
pasta structures to determine whether the phase transitions obtained
in nuclear matter~\cite{dorso2018_arxiv} survive in
neutron star matter, and what is the behavior of the symmetry
energy within the pasta.

\subsection{\label{sym_energy}Symmetric neutron star matter}

We first study the case of symmetric neutron star matter, i.e with 
$x = z/A= 0.5$, focusing on the caloric curve and the binding 
energy.

\subsubsection{\label{subsec:sym_energy_1}The caloric curve}

For starters we calculate the internal energy of symmetric neutron star 
matter. This involves averaging the kinetic and potential energy of each 
nucleon in the system. \\

To detect possible phase changes it is customary to calculate the caloric 
curve, which plots the relationship between the internal energy and the 
temperature. This curve indicates the amount of heating produced as 
a function of the energy added to the system. During a phase transition, 
however, the addition of energy does not produce a change in temperature 
because the energy is not used to heat the system but to break bonds 
melting the system or liberating liquid molecules. Thus, changes in the 
slope of the $E-T$ curve can be used as indicators of phase transitions.\\

In our case it is convenient to remember that, as seen in 
Ref.~\cite{dorso2018_arxiv}, nuclear matter maintains a liquid-like structure 
for temperatures larger than about 2.0 MeV, temperature at which it freezes 
into a pasta structure composed of nucleons arranged amorphously (still 
like a liquid).  And at around $T=0.5$ MeV it transforms into a pasta 
with nucleons frozen much like in crystalline structures.\\

\begin{figure}
\begin{center}
   \includegraphics[width=\columnwidth]{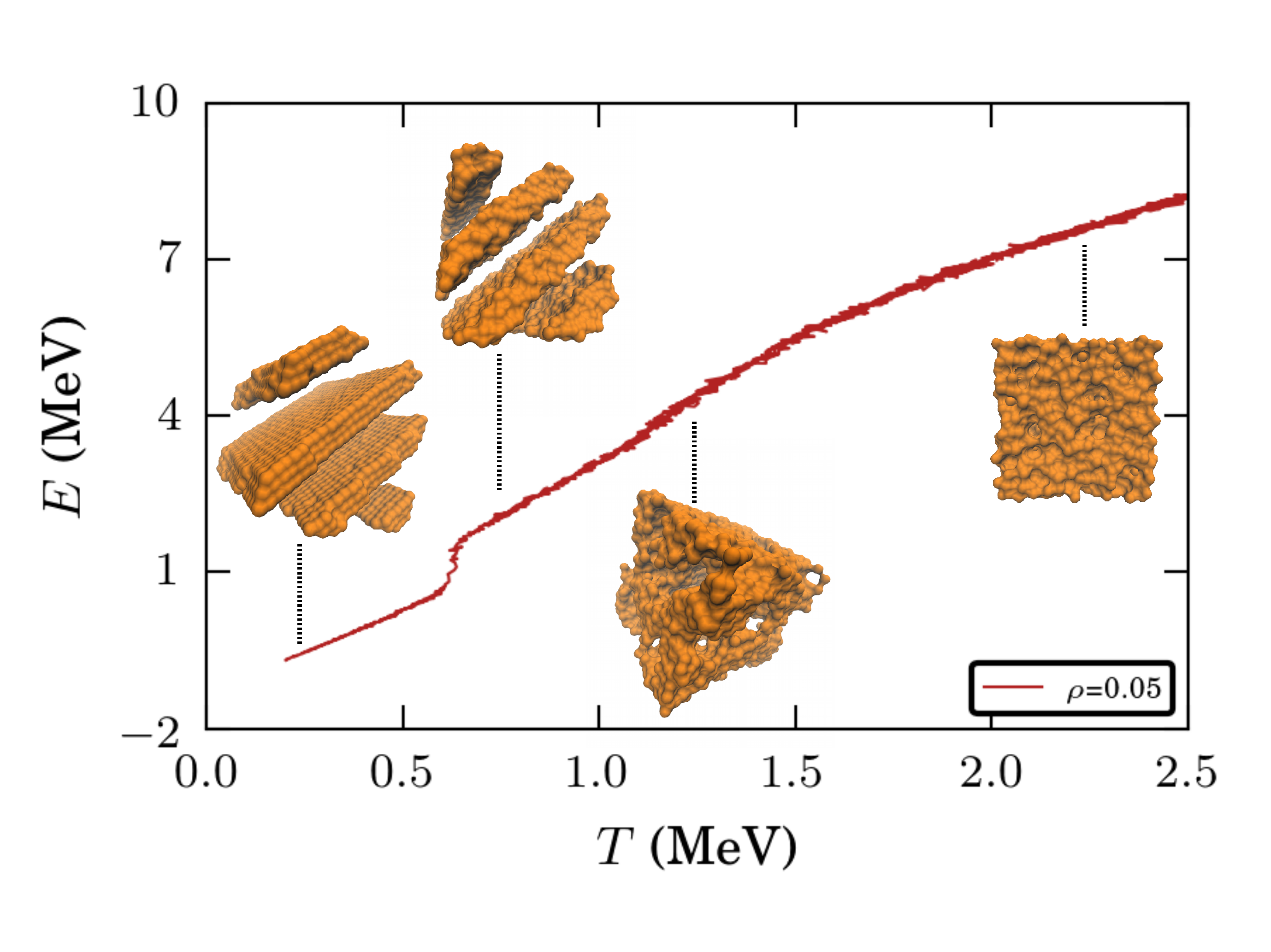}
\caption{(color online) Internal energy per nucleon for symmetric neutron star 
matter ($x=0.5$) as a function of the bath temperature, for 
$\rho=0.05\,$fm$^{-3}$. }\label{fig:eos_0}
\end{center}
\end{figure}

\begin{figure}[!htbp]
\begin{center}
   \includegraphics[width=\columnwidth]{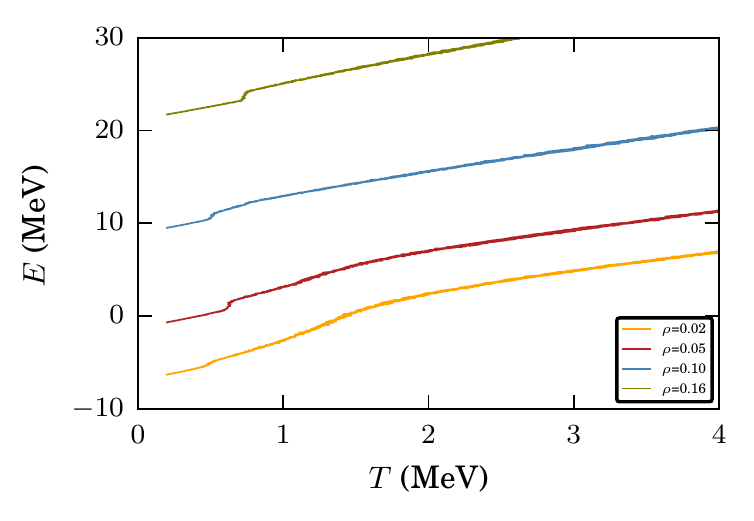}
\caption{\label{fig:eos_1} (color online) Internal energy per nucleon for 
symmetric neutron star matter ($x=0.5$) as a function of the bath temperature. 
The mean density for each profile is indicated in the insert.}
\end{center}
\end{figure}

For neutron star matter the results are somewhat different. Figure~\ref{fig:eos_0} 
presents the case of $\rho=0.05$ fm$^{-3}$, and Figure~\ref{fig:eos_1} 
shows the curves for densities $\rho=$0.02, 0.05, 0.10 and 0.16 fm$^{-3}$ 
in the extended temperature range of 0.2 to 4 MeV. In both figures the total 
number of nucleons in the primary cell was $N=4000$ interacting through the New 
Medium model (c.f. Section~\ref{NSP}). \\

As it can be seen more clearly in Fig.~\ref{fig:eos_0}, the slope of the 
internal energy exhibits a change at around $T\simeq 1.5\,$MeV, and 
a sharper change at $T\simeq 0.5\,$MeV. Although these jumps are not 
as pronounced as those found for NM, they happen at around 
the same temperatures and can be taken as the same type of transitions.\\

We claim that the smooth change at $T\simeq 1.5\,$MeV signals 
the onset of the topological phase transition, (\textit{i.e.} a bubble 
appears). A difference with NM is that this change in slope appears more 
pronounced for smaller densities ($\rho \leq \, 0.05\,$fm$^{-3}$), and tends to 
disappear for larger densities. As this result is different than the NM case, we 
believe it is due to the presence of the electron gas. \\

On the other hand, the sharper jump observed at $T\simeq 0.5\,$MeV 
both in Figs.~\ref{fig:eos_0} and ~\ref{fig:eos_1} is maintained 
throughout the density range studied and, in fact, becomes more 
pronounced for larger densities. We believe this discontinuity in the 
derivative of the $E-T$ curve signals the amorphous pasta-crystalline 
pasta change already identified in Ref.~\cite{dorso2018_arxiv} for NM  
with the Pandharipande Medium model.\\

\begin{figure}
\begin{center}
   \includegraphics[width=\columnwidth]{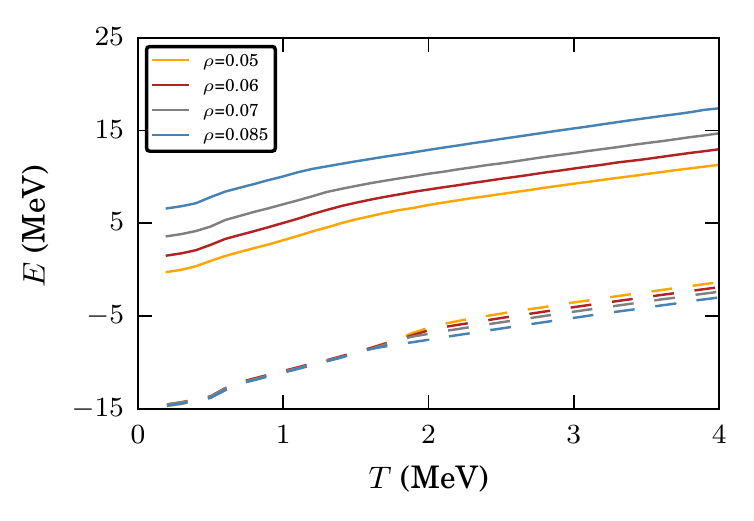}
\caption{(color online) Internal energy per nucleon for symmetric neutron
star matter ($x=0.5$) as a function of temperature. The mean density for each 
profile is indicated in the inset in fm$^{-3}$. The dashed lines correspond to 
the Pandharipande Medium model (nuclear matter), while the continuous lines 
correspond to the New Medium potential with the screened Coulomb 
potential.}\label{fig:eos_3}
\end{center}
\end{figure}

The noticeable differences between the caloric curve of nuclear matter and that 
of neutron star matter are shown in Fig.~\ref{fig:eos_3}. The figure compares 
the NSM results (continuous lines) against the NM curves (dashed lines) for the 
cases of $\rho=0.05$, 0.06, 0.07 and 0.085 fm$^{-3}$ and temperatures from 0.2 
to 4 MeV. The NSM lines remain separated and parallel throughout the range of 
explored temperatures, whereas the NM curves merge with one another at low 
temperatures. Noteworthy, the NM highest density curve ($\rho=0.085\,$fm$^{-3}$) 
corresponds to the lowest energy, while the opposite is true for NSM where the 
higher density curve has the higher energy of all the curves shown.

\subsubsection{\label{subsec:sym_energy_2}The energy}

When plotted against the density, the energy can be a useful 
indicator of the state of the system. As seen in the case of 
NM~\cite{dorso2018_arxiv}, the $E-\rho$ curve can signal the saturation density 
(\textit{i.e.} the minimum of the curve), bound and unbound states (positive or 
negative energies), the range of densities of the liquid phase (span of the 
``$\cup$'' shape), and changes of phase, among others. For the case of NSM, 
however, the situation is totally different.\\

\begin{figure}[!htbp]
\begin{center}
   \includegraphics[width=\columnwidth]{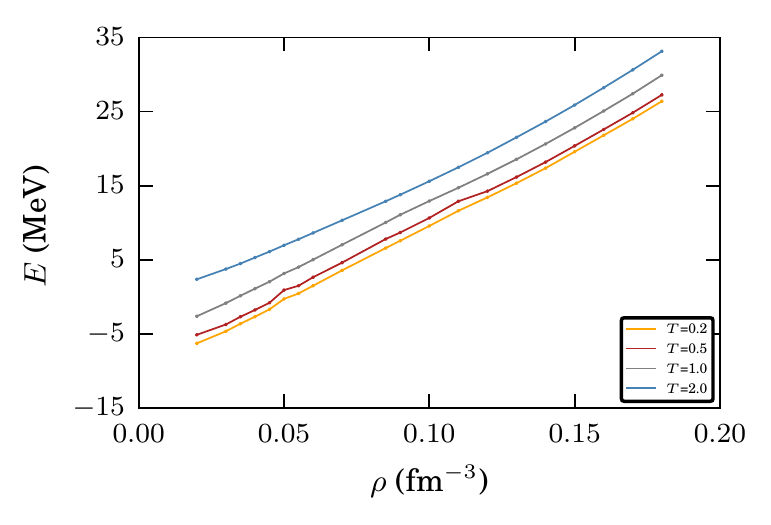}
\caption{\label{fig:eos_2} (color online) Internal energy per nucleon for 
symmetric neutron star matter ($x=0.5$) as a function of the mean density at 
fixed temperatures. The corresponding bath temperature for each is indicated in 
the insert.}
\end{center}
\end{figure}

Fig.~\ref{fig:eos_2} shows the energy as a function of the average
density at $T = 0.1$, $0.2$, $1.0$ and $2.0$ MeV, as obtained from the CMD 
simulations; these results, interestingly, differ substantially from the
behavior exhibited by symmetric nuclear matter. As shown in the four 
 lines, there is no absolute minima and, thus, NSM at these temperatures and 
densities does not have an equilibrium point nor a saturation density; this is 
different than the $\cup$ shape that what was found for NM. Likewise, at a 
difference from NM, for most of the densities the NSM system is unbound, except 
for $T\leq 1.0\,$MeV and $\rho\leq 0.05\,$fm$^{-3}$. \\

The difference between nuclear matter and neutron star matter is that for NM 
there is no embedding electron gas, whereas for NSM the effect of the cloud of 
electrons introduces the short range Coulomb screening. It is quite interesting 
that because the exponential cut-off is of the order of 20 fm, small aggregates 
of nucleons resemble nuclei quite closely.\\

\subsubsection{\label{subsec:sym_topology}The Minkowski functionals}

A first insight to the topology associated to the structures 
formed may be given by the Minkowski functionals. As explained before, the 
Minkowski functionals~\cite{michielsen} use the volume, surface area, Euler 
characteristic $\chi$, and integral mean curvature B to characterize the size, 
shape and connectivity of spatial structures formed by the nucleons. The 
calculation of the Minkowski functionals requires the binning of nucleons into 
``voxels'' which have to be of the proper size (see, e.g., the Appendix of 
Ref.~\cite{dorso2018_arxiv} for the procedure to use). In this work cubic voxels 
of $d=2.35\,$fm of length were used, and the Euler functional $\chi$ was 
computed according to Eq.~(\ref{eq:chi}). \\

\begin{figure}
\begin{center}
   \includegraphics[width=\columnwidth]{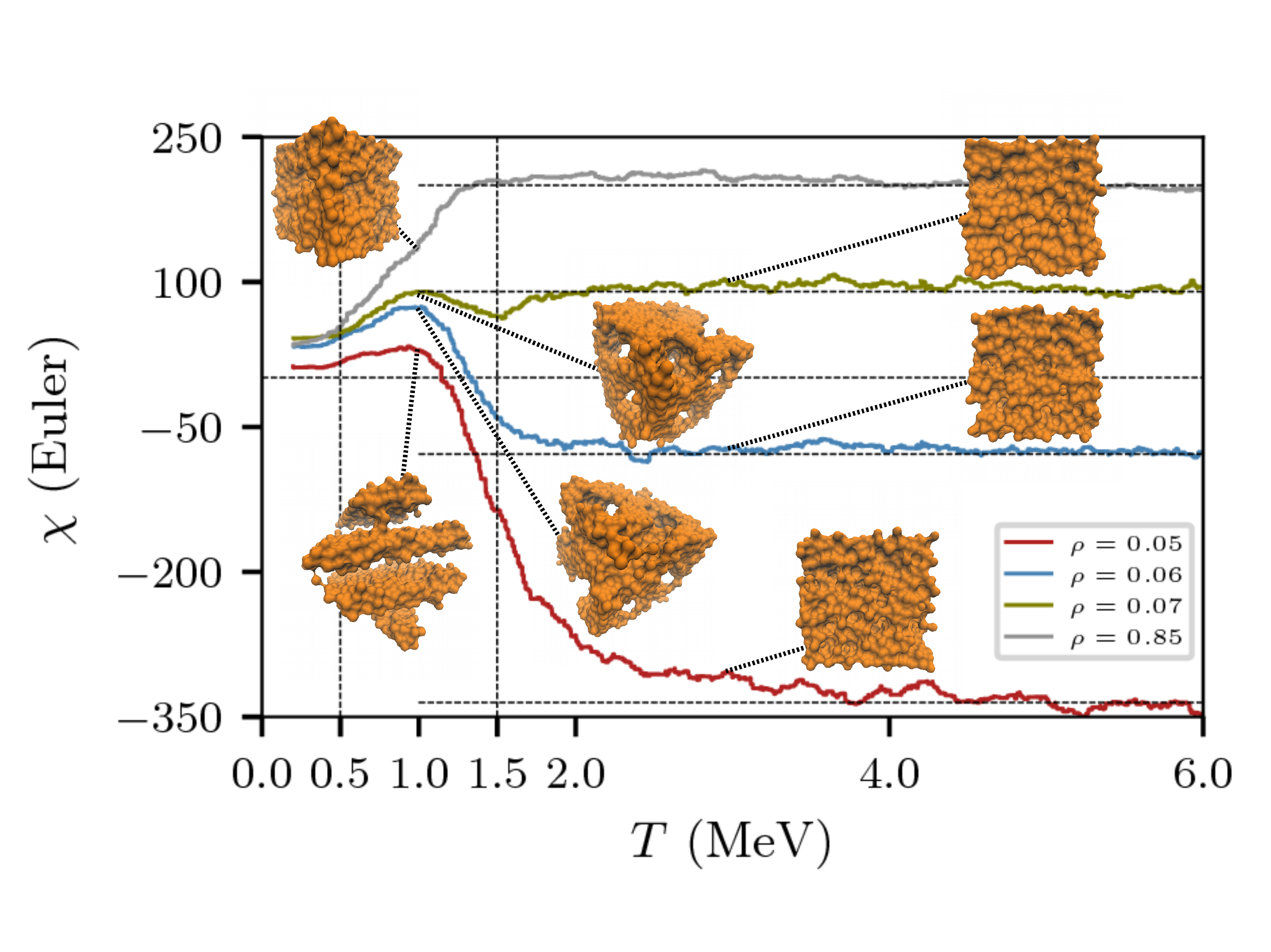}
\caption{(color online) The Euler functional $\chi$ obtained for isospin 
symmetric ($x=0.5$)  NSM systems as a function of temperature. The systems had 
the densities shown in the inset (in fm$^{-3}$) and were composed by $N=4000$ 
nucleons interacting through the New Medium model. The corresponding binning 
distance is $d=2.35\,$fm. The data has been smoothed with a moving average 
procedure.}\label{fig:minkowski_rho}
\end{center}
\end{figure}

Figure~\ref{fig:minkowski_rho} shows the Euler functional $\chi$ as a function 
of the the temperature. Although the $\chi$ may somewhat depend on the binning 
distance of the simulation cell, the exhibited patterns are qualitatively 
meaningful. Several features are worth mentioning. 

$\chi$ has a clear change of behavior at $T\simeq 1\,$MeV. Although this happens 
at a lower temperature than its NM counterpart (see Fig. 16 in 
Ref.~\cite{dorso2018_arxiv}), it also appears to be associated to the early 
stage of the pasta formation.  Furthermore, the $\chi$ values for the examined 
densities almost join into a single pattern for $T\leq 0.5\,$MeV, much like 
those of NM at the same temperature but less pronounced; this could also be 
related to the change of slope found in the caloric curve at $T\simeq 0.5$ 
MeV.\\

A look at the sign of $\chi$ can yield information about the morphology of the 
structure. Figure~\ref{fig:minkowski_rho} indicates that the lower density 
systems attain negative values of $\chi$ at, say, $T>1\,$MeV. Since values of 
$\chi < 0$ corresponds to cases where the number of tunnels overcome the number 
of voids and isolated regions, see Eq.~(\ref{eq:chi}), it seems that low density 
configurations tend to be more cavity-like (more tunnels), but as density 
increases tunnels fill up yielding more compact structures.\\

In summary, and comparing to NM, the introduction of the coulomb 
screened potential appears to smooth out the phase transitions. The Euler 
functional $\chi$ experiences a somewhat smooth change along $1-2\,$MeV, in 
correspondence with the energy changes mentioned in 
Section~\ref{subsec:sym_energy_1}. Thus, the ``pasta'' forming process may be 
located at this temperature range, in a similar fashion as in nuclear matter 
systems (see Ref.~\cite{dorso2018_arxiv}). At temperatures below $T\simeq 
0.5\,$MeV the (inner) nucleons in the ``pasta'' structure freeze into a solid 
state. 

\subsubsection{\label{subsec:rdf}The radial distribution function}

\begin{figure}
\begin{center}
   \includegraphics[width=\columnwidth]{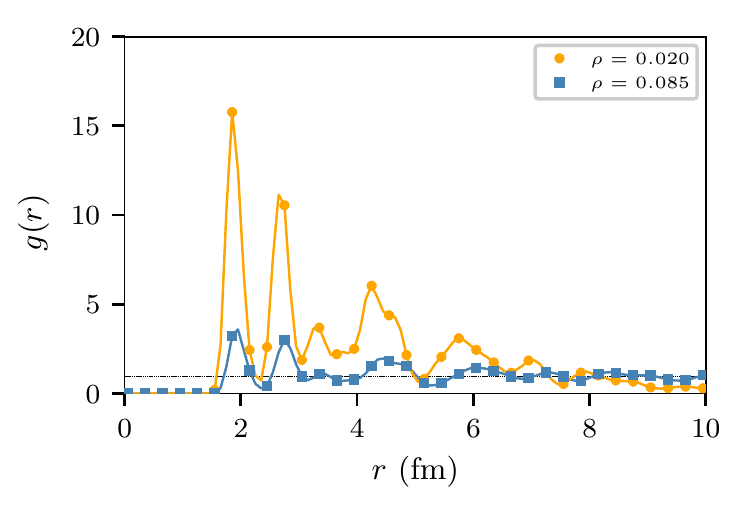}
\caption{(color online)  Radial distribution function $g(r)$ for nucleons 
corresponding to a symmetric neutron star matter system of $N=4000$ nucleons and 
$T=0.2\,$MeV. The system mean density is indicated in the inset in fm$^{-3}$. 
The horizontal line at $1$ corresponds to the asymptotic limit expected for 
infinite systems.}\label{fig:gr_rho02_rho085_all}
\end{center}
\end{figure}

The radial distribution function $g(r)$ can be used to explore the phase 
transformations detected by the caloric curve by looking at the average 
distribution of nearest neighbors. Fig.~\ref{fig:gr_rho02_rho085_all} shows the 
$g(r)$ for two systems with $N=4000$ nucleons interacting through the New Medium 
model at two sub-saturation densities. The $\rho=0.02$ profile resembles a 
more regularly distributed system, with more pronounced nearest neighbor peaks, 
as compared to the $\rho=0.085$ profile. Undoubtedly, the lower density case 
corresponds to a more crystalline phase than the high density which appears less 
structured and, looking at Figure~\ref{fig:eos_3}, we see that the highest 
internal energy is associated to a more regular distributions of nucleons within 
the pasta regime.\\

\begin{figure*}[!htbp]
\centering
\subfloat[$\rho=0.02$\label{fig:frame_rho01_x05}]{
\includegraphics[width=0.66\columnwidth]
{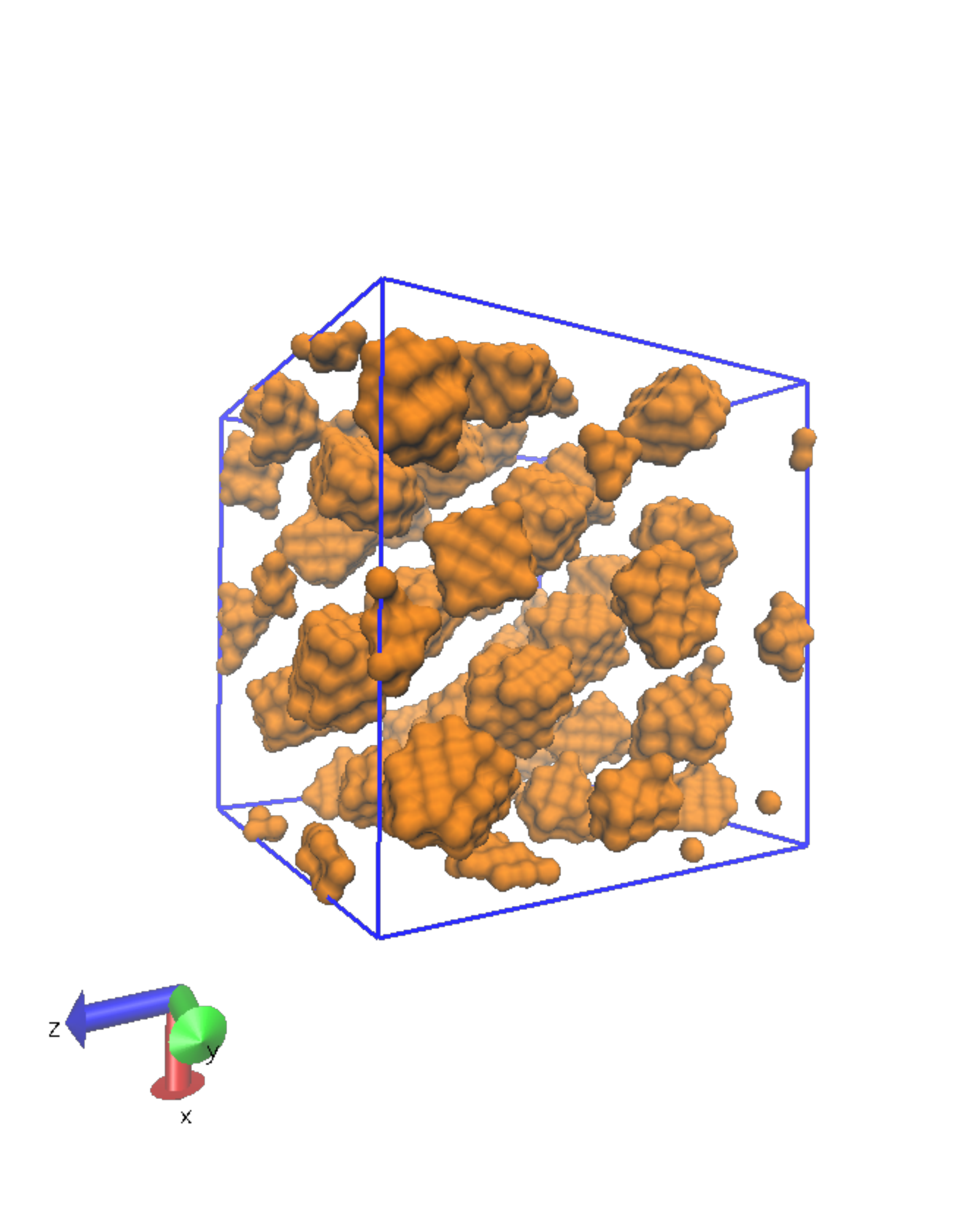}
}
\subfloat[$\rho=0.04$\label{fig:frame_rho04_x05}]{
\includegraphics[width=0.66\columnwidth]
{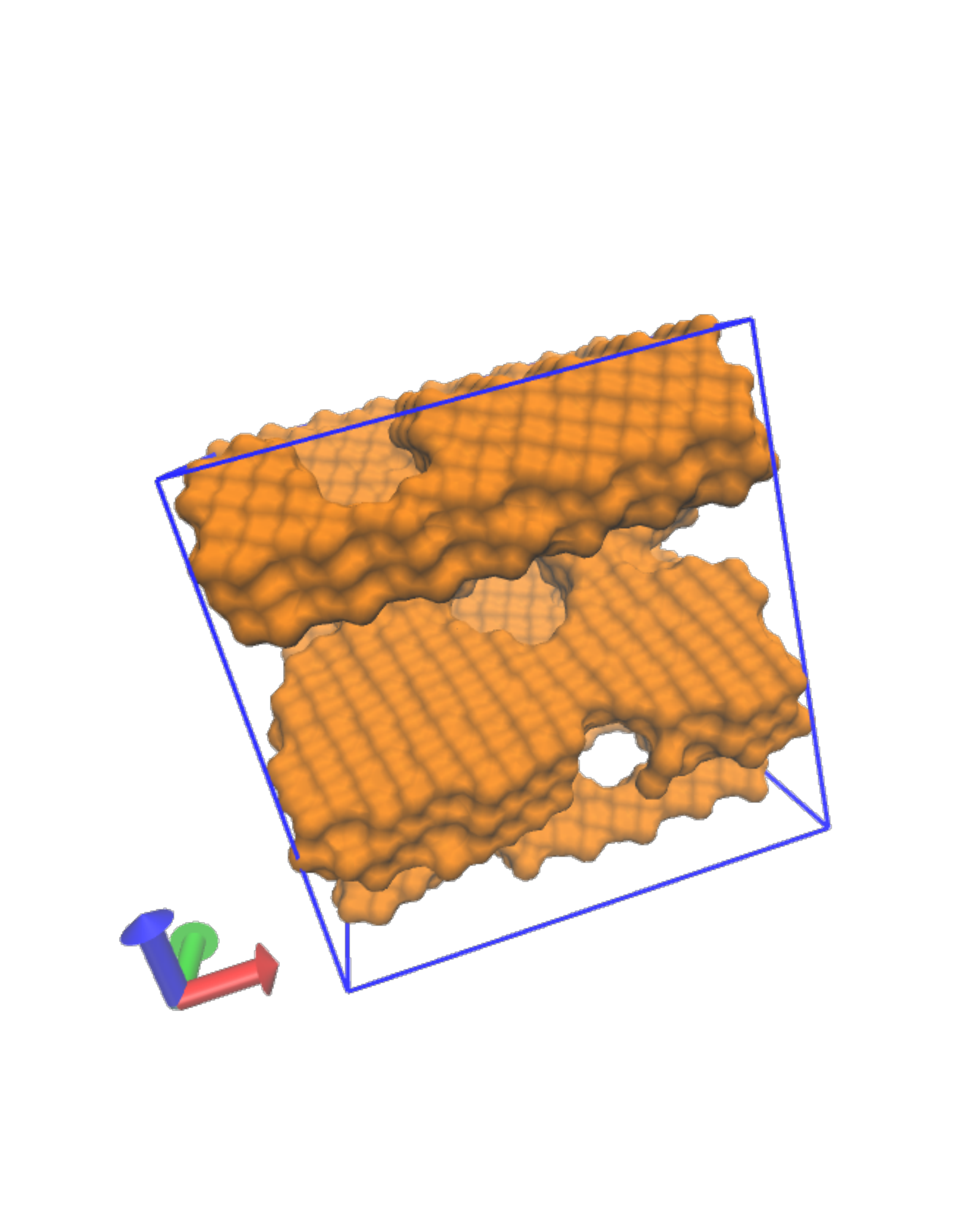}
}
\subfloat[$\rho=0.085$\label{fig:frame_rho085_x05_a}]{
\includegraphics[width=0.66\columnwidth]
{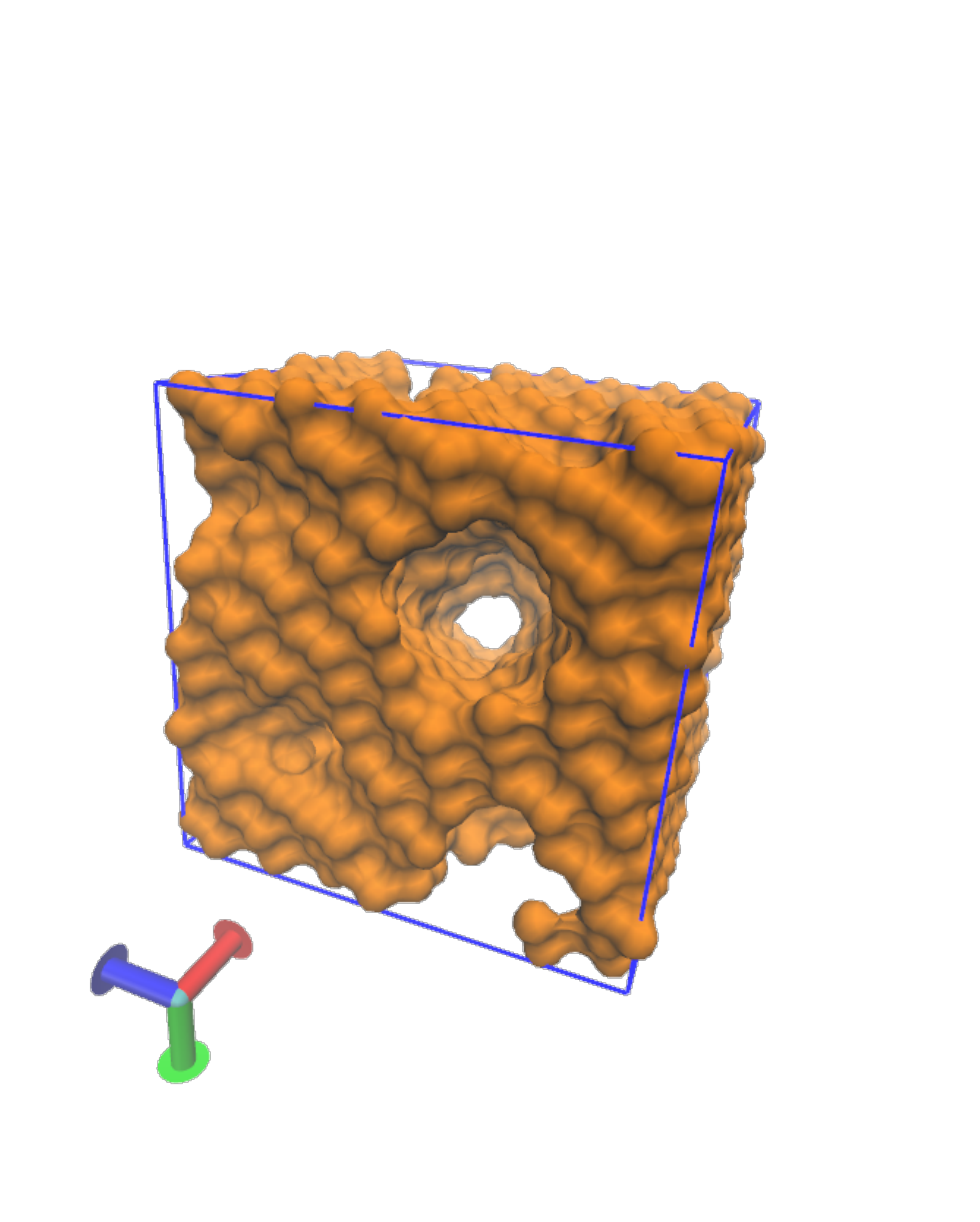}
}
\caption{\label{fig:frame_rho01_rho04_rho085_sym} (color online) Surface 
representation for the protons at the indicated densities. The snapshot was 
taken at $T=0.2$ and $x=0.5$ for a system of $4000$ nucleons.  }
\end{figure*}

The correlation of $g(r)$ to the pasta morphology can be observed by plotting 
the spatial distribution of the protons. 
Fig.~\ref{fig:frame_rho01_rho04_rho085_sym} shows the proton arrangements for 
three density situations. It can be checked from 
Fig.~\ref{fig:frame_rho085_x05_a} that the $\rho=0.085$ situation is highly
homogeneous inside the occupied regions, although hollow spaces (of protons)
exist. These occupied regions split into smaller pieces, according to  
Figs.~\ref{fig:frame_rho04_x05} and \ref{fig:frame_rho01_x05}, forming
\textit{lasagnas} and \textit{gnocchis}, respectively. Thus, as the density
diminishes (and the structures break into smaller pieces), the internal energy
decreases (see fig.~\ref{fig:eos_2}).\\

The connection to $g(r)$ is as follows. At the lowest explored density 
($\rho=0.02$ fm$^{-3}$) of Figure~\ref{fig:gr_rho02_rho085_all} $g(r)$ shows 
sharp peaks at $r\leq6\,$fm $^{-3}$ and almost no correlation ($g(r) \approx 0$) 
for $r\geq8\,$fm$^{-3}$. This indicates that strong correlations exist between 
neighbors belonging to the same \textit{gnocchi}, and no correlation at large 
separation distances between \textit{gnocchis} (see 
Fig.~\ref{fig:frame_rho01_x05}). For the larger density, $\rho=0.085$ fm$^{-3}$, 
$g(r)$ tends to $1$ as expected for homogeneous systems (c.f. 
\ref{fig:frame_rho085_x05_a}).\\

We may summarize Section~\ref{sym_energy} as follows. The internal energy for
symmetric neutron star matter increases monotonically for either increasing 
temperatures and densities; it does not appear to have a saturation point.  
Since these effects do not occur for nuclear matter, it can be said that the 
Coulomb potential is responsible for biasing the energy upward, as can be seen 
clearly by comparing Fig.~\ref{fig:eos_1} to the nuclear matter case. Likewise, 
the introduction of the Coulomb potential affects the system morphology; as the 
density diminishes, the pasta structures split into smaller structures.

\subsection{\label{subsec:asym}Non-symmetric neutron star matter}

\subsubsection{\label{subsec:asym_energy}The internal energy}

As a second step in our study of NSM we explored the behavior of the internal 
energy in systems with decreasing fractions of protons. We investigated 
fractions from $x = 0.1$, 0.2, 0.3 and 0.4 to mimic the proton fractions 
appearing in neutron crusts. All simulations consist of systems with $N=4000$ 
nucleons under periodic boundary conditions and interacting through the New 
Medium potential.\\

\begin{figure*}[!htbp]
\centering
\subfloat[$\rho=0.04$\label{fig:eos_4}]{
\includegraphics[width=\columnwidth]
{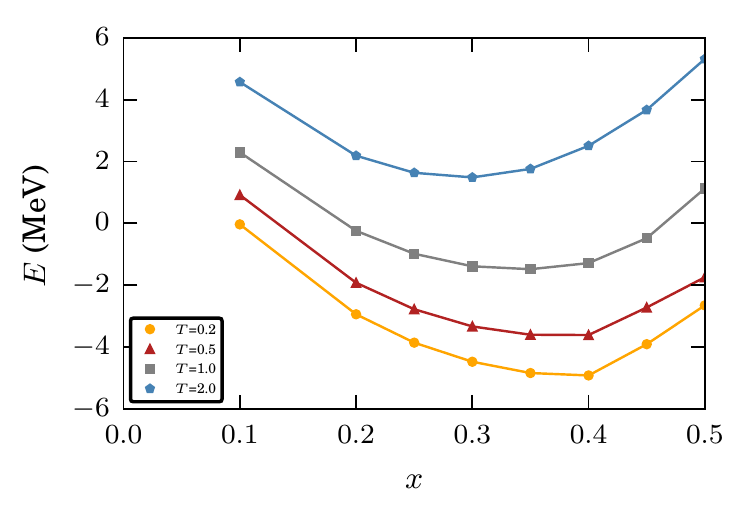}
}
\subfloat[$\rho=0.085$\label{fig:eos_5}]{
\includegraphics[width=\columnwidth]
{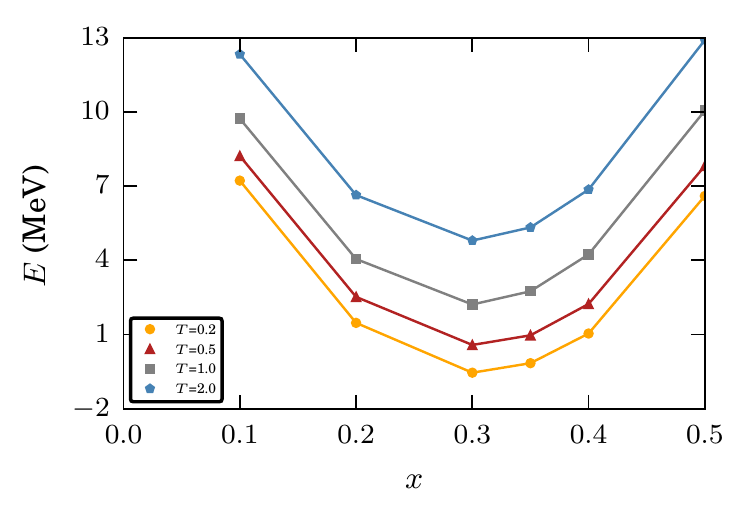}
}
\caption{\label{fig:eos_x} (color online)
Internal energy per nucleon for neutron star matter systems with $N=4000$ 
nucleons. All the curves correspond to the New Medium model (see 
Section~\ref{NSP}). (a) Mean density $\rho=0.04$. (b) Mean density $\rho=0.085$. 
}
\end{figure*}

Figures~\ref{fig:eos_4} and \ref{fig:eos_5} show the isothermal energies as a 
function of $x$ for systems with densities $\rho=0.04\,$fm$^{-3}$ and 
$\rho=0.085\,$fm$^{-3}$. Each of the curves exhibits a $\cup$ shape indicating 
the existence of a minimum of the energy at a certain value of $x$. At 
$\rho=0.04$ fm$^{-3}$ the observed minima appears to shift from $x=0.4$ at 
$T=0.2$ MeV to $x=0.3$ at $T=2.0$ MeV, while at $\rho=0.085$ fm$^{-3}$ the 
minima remain at $x=0.3$ at all temperatures. This finding indicates that 
systems with the freedom of exchanging their content of neutrons and protons 
would favor a specific isospin ratio $x$ depending on the local temperature and 
density of the system.\\

\begin{figure*}[!htbp]
\centering
\subfloat[$T=0.2$\label{fig:eos_6}]{
\includegraphics[width=\columnwidth]
{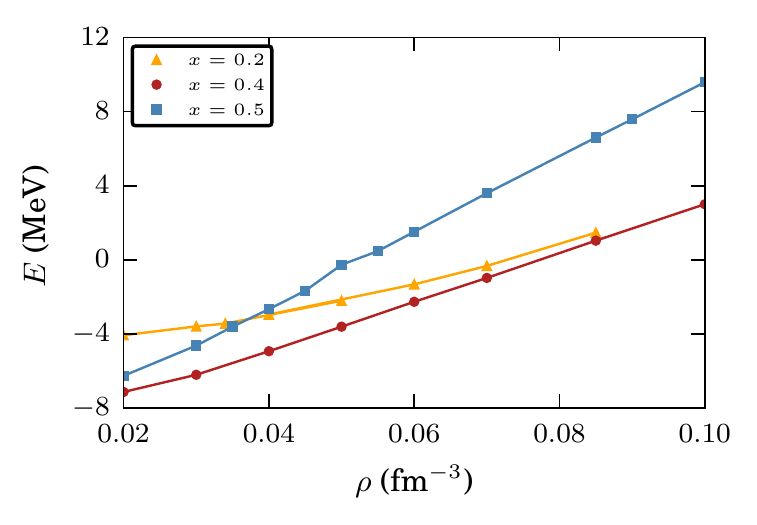}
}
\subfloat[$T=1.0$\label{fig:eos_7}]{
\includegraphics[width=\columnwidth]
{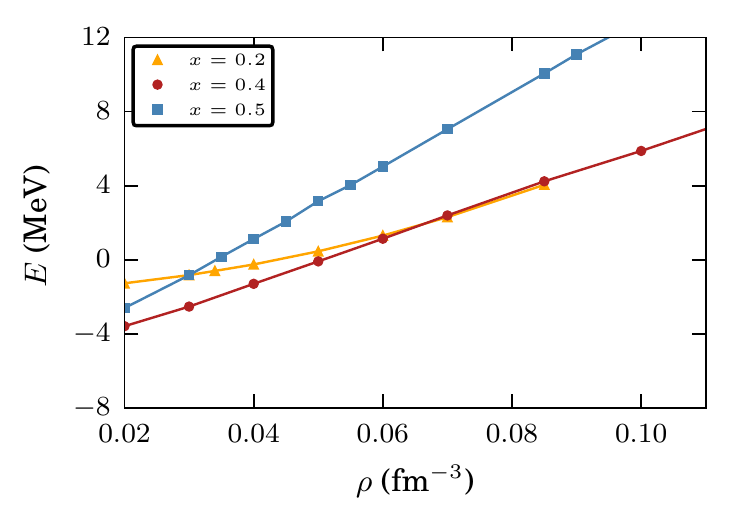}
}
\caption{\label{fig:eos_rho} (color online) Internal energy per nucleon for 
neutron star matter systems of $N=4000$ nucleons at three proton fractions and 
interacting through the New Medium model. (a) Temperature $T=0.2\,$MeV. (b) 
Temperature $T=1.0\,$MeV. }
\end{figure*}

Complementary information can be obtained from the energy dependence of the 
energy isotherms.  Figure~\ref{fig:eos_rho} shows the corresponding curves for 
$x=0.2$, 0.4 and 0.5. It is interesting to notice that while the $x=0.4$ and 0.5 
curves appear to have similar monotonically decreasing behaviors at all 
densities, the $x=0.2$ case deviates at low densities. As we will see next, this 
is due to the $\cup$ shape $x$ dependence of the energy seen in 
Figures~\ref{fig:eos_4} and \ref{fig:eos_5}.\\

Specifically, Panel~\ref{fig:eos_6} shows the $T=0.2$ MeV case and, if focusing 
at $\rho = 0.04$ fm$^{-3}$, we see that $E(x=0.4)<E(x=0.2)\approx E(x=0.5)$, as 
can be expected by looking at the points at $x$ =0.2, 0.4 and 0.5 of the $T=0.2$ 
MeV isotherm in Figure~\ref{fig:eos_4}. Similarly, Panel~\ref{fig:eos_7} can be 
understood by comparing to the $T=1.0$ MeV isotherm in Figure~\ref{fig:eos_5}. 
One can see, then, that the diverging behavior of the $x=0.2$ energy-density 
curves at low densities in Figure~\ref{fig:eos_rho} is simply explained by the 
$\cup$ dependence of the energy on the isospin content $x$.\\

\subsubsection{\label{subsec:asym_topology}The associated topology}

\begin{figure*}[!htbp]
\centering
\subfloat[$\rho=0.04$\label{fig:minkowski_asym_rho04}]{
\includegraphics[width=1.05\columnwidth]
{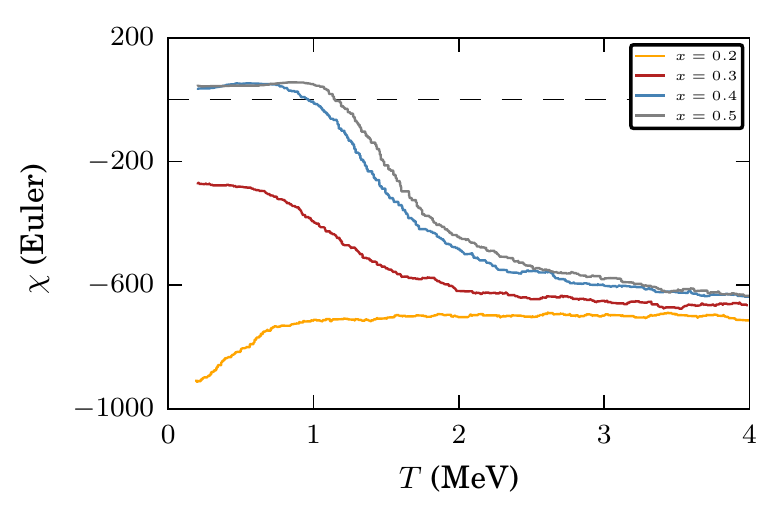}
}
\subfloat[$\rho=0.085$\label{fig:minkowski_asym_rho085}]{
\includegraphics[width=\columnwidth]
{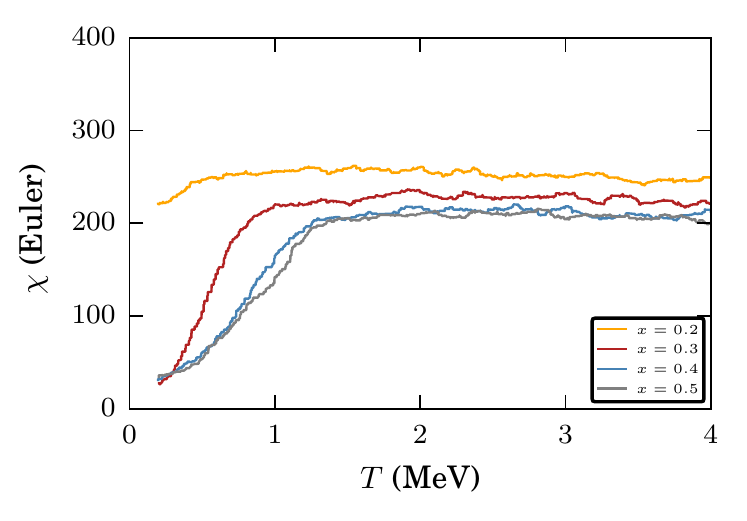}
}
\caption{\label{fig:minkowski_asym} (color online) The Euler functional $\chi$ 
as a function of temperature for non-symmetric New Medium model. The 
corresponding proton fractions $x$ are indicated in the inset. The total number 
of nucleons is $N=4000$. The binning distance is $d=2.35\,$fm. The horizontal 
line corresponds to the null level.  (a) The system mean density is 
$\rho=0.04\,$fm$^{-3}$. (b) The system mean density is $\rho=0.085\,$fm$^{-3}$. 
The data has been smoothed with a moving average procedure. }
\end{figure*}

Following the footsteps of the case of symmetric NSM, we proceed to examine the 
Euler functionals $\chi$ and the radial distribution functions $g(r)$ for the 
non-symmetric systems. Fig.~\ref{fig:minkowski_asym} shows the $\chi$ for four 
representative proton fractions at the densities $\rho=0.04\,$fm$^{-3}$ and 
$\rho=0.085\,$fm$^{-3}$.\\

In general terms, $\chi$ indicates changes of structure at around $T\approx1$ 
MeV. Figure~\ref{fig:minkowski_asym_rho04} shows that at the lower density 
($\rho=0.04$ fm$^{-3}$) and at temperatures hotter than $T\ge2$ MeV, the system 
appears dominated by voids and tunnels independent of the isospin content, while 
at lower temperatures ($T\le 1$ MeV) more compact objects appear at all $x$, 
except for $x=0.2$ which disperses even more.  
Figure~\ref{fig:minkowski_asym_rho085} shows that at the higher density 
($\rho=0.085$ fm$^{-3}$) the system attains a more compact structure at all 
isospin contents and for $T\ge 1$ MeV, becoming less pronounced at lower 
temperatures ($T\le 1$ MeV).\\

\begin{figure*}[!htbp]
\centering
\subfloat[$x=0.2$\label{fig:frame_rho085_x02}]{
\includegraphics[width=0.66\columnwidth]
{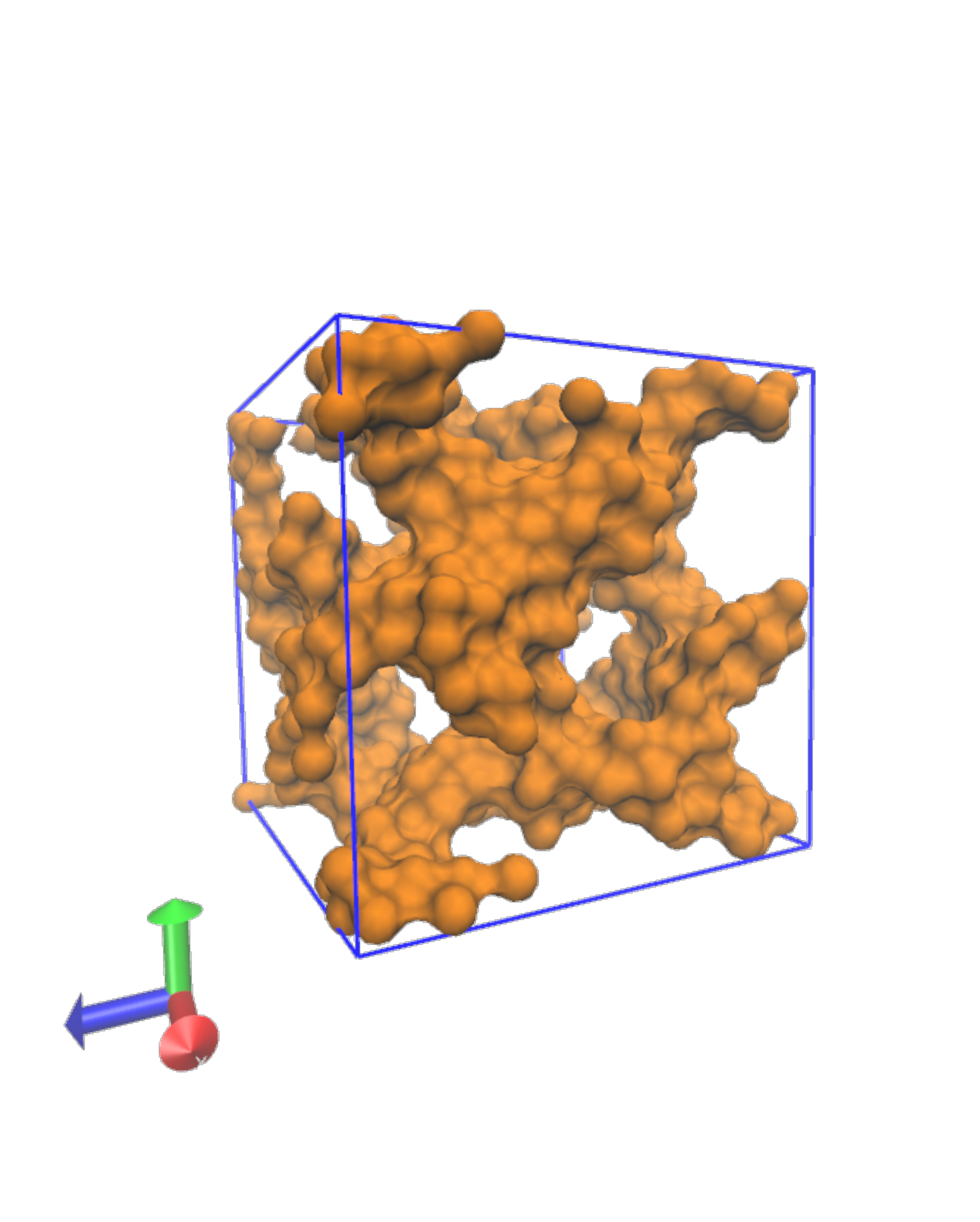}
}
\subfloat[$x=0.4$\label{fig:frame_rho085_x04}]{
\includegraphics[width=0.66\columnwidth]
{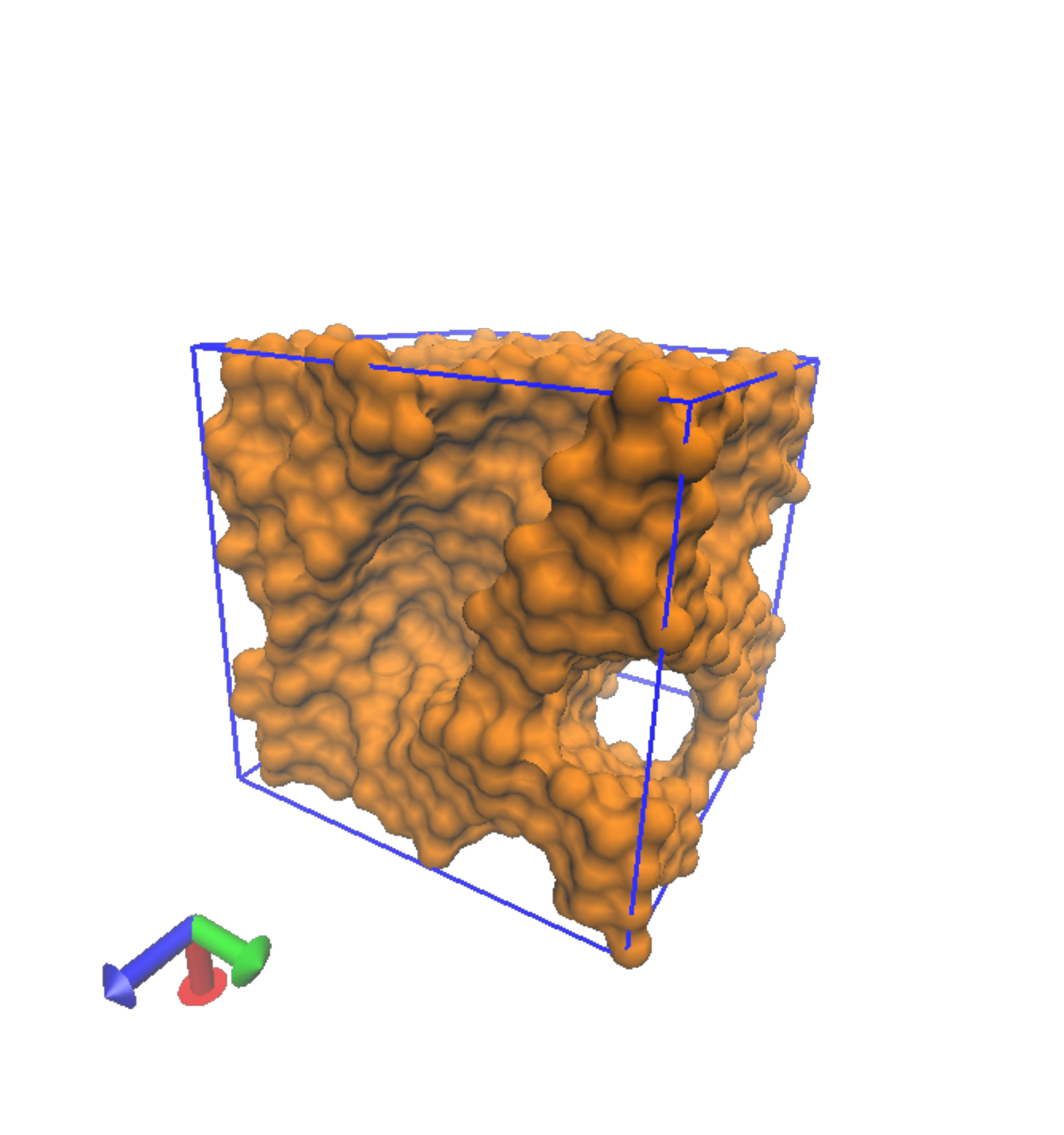}
}
\subfloat[$x=0.5$\label{fig:frame_rho085_x05}]{
\includegraphics[width=0.66\columnwidth]
{./lammps_rho085_x05_protons_surf_star.pdf}
}
\caption{\label{fig:frame_rho085_asym} (color online)  Surface representation 
for the protons at the indicated proton fractions. The snapshot was taken at 
$T=0.2\,$MeV and $\rho=0.085\,$fm$^{-3}$ for a system of $4000$ nucleons under 
periodic boundary conditions interacting through the New Medium Model plus the 
screened Coulomb potential.  }
\end{figure*}

Figure~\ref{fig:frame_rho085_asym} shows the space position of the protons for 
the cases of $x=0.2$ (a), $x=0.4$ (b), and $x=0.5$ (c) at $T=0.2$ and 
$\rho=0.085$ fm$^{-3}$. Although not directly comparable to the values of $\chi$ 
obtained in Figure~\ref{fig:minkowski_asym_rho085} (recall that the computation 
of $\chi$ includes both neutrons and protons), the three structures in 
Figures~\ref{fig:frame_rho085_asym} correspond to the left end points ($T=0.2$ 
MeV) of the $\chi$ curves in Figure~\ref{fig:minkowski_asym_rho085}; notice that 
no major changes in the percentage of void to filled volumes are observed as it 
is difficult to appreciate such changes in compact structures.\\

\begin{figure*}[!htbp]
\centering
\subfloat[$x=0.1$\label{fig:frame_rho04_x01}]{
\includegraphics[width=0.66\columnwidth]
{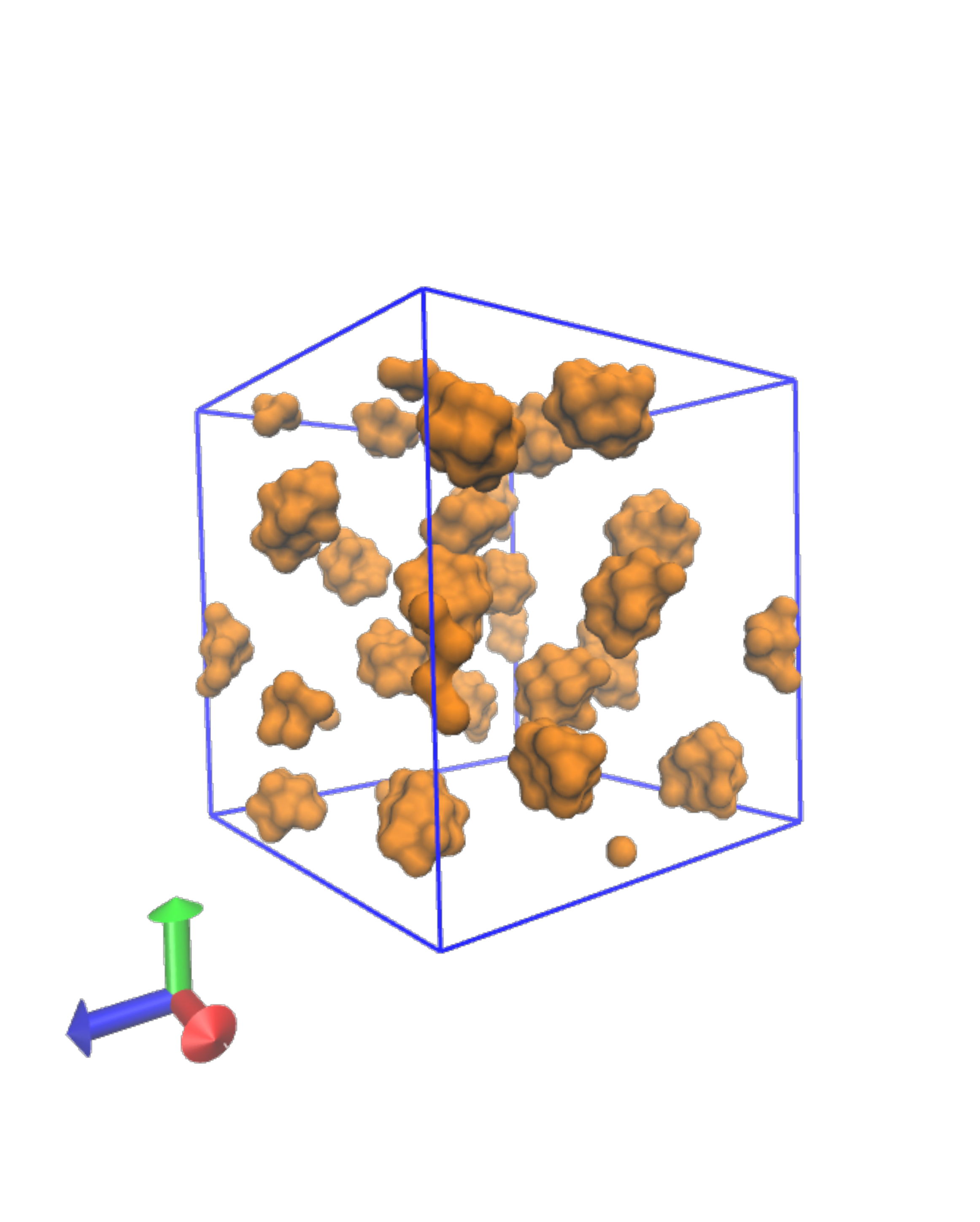}
}
\subfloat[$x=0.2$\label{fig:frame_rho04_x02}]{
\includegraphics[width=0.66\columnwidth]
{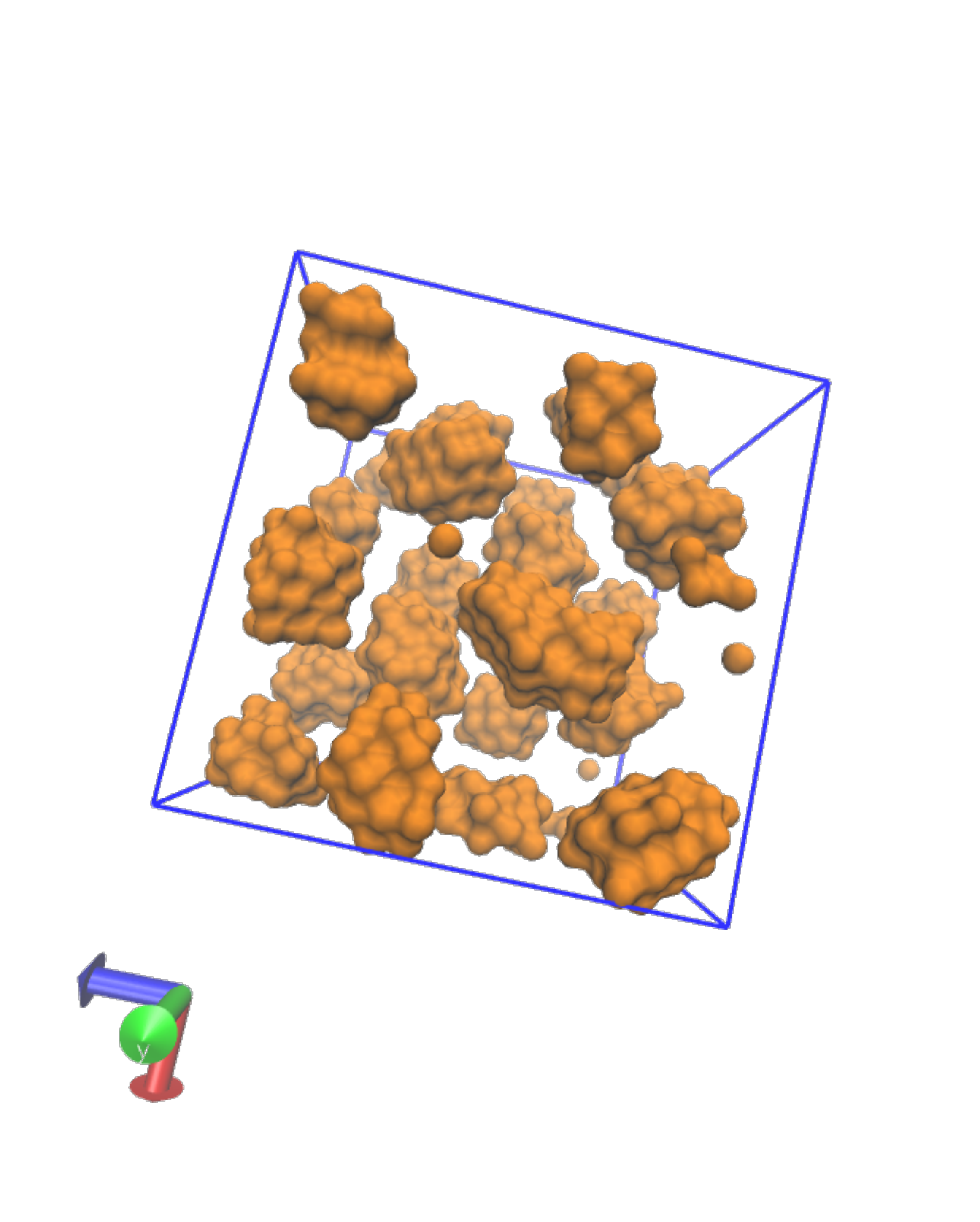}
}
\subfloat[$x=0.3$\label{fig:frame_rho04_x03}]{
\includegraphics[width=0.66\columnwidth]
{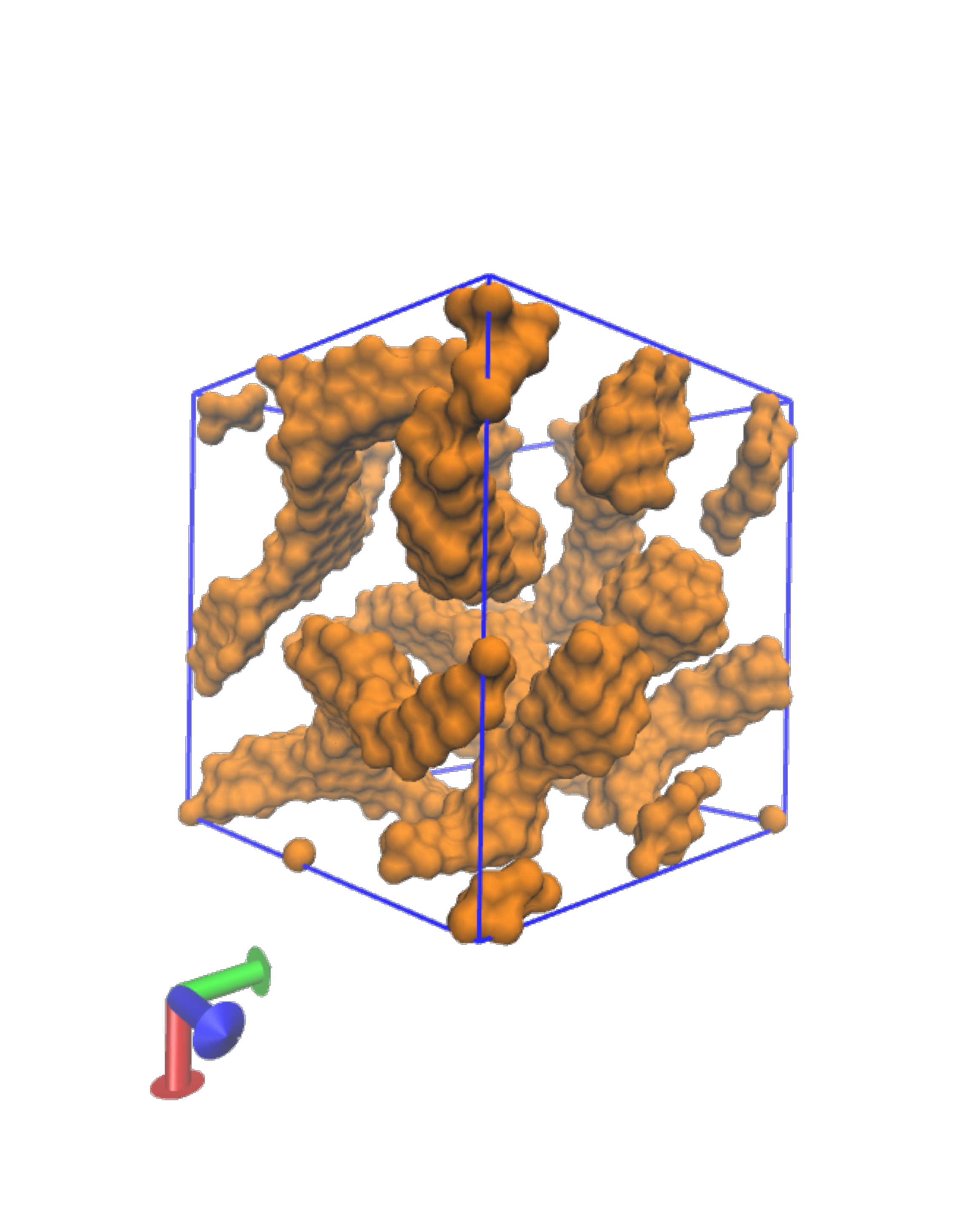}
} \\
\subfloat[$x=0.4$\label{fig:frame_rho04_x04}]{
\includegraphics[width=0.66\columnwidth]
{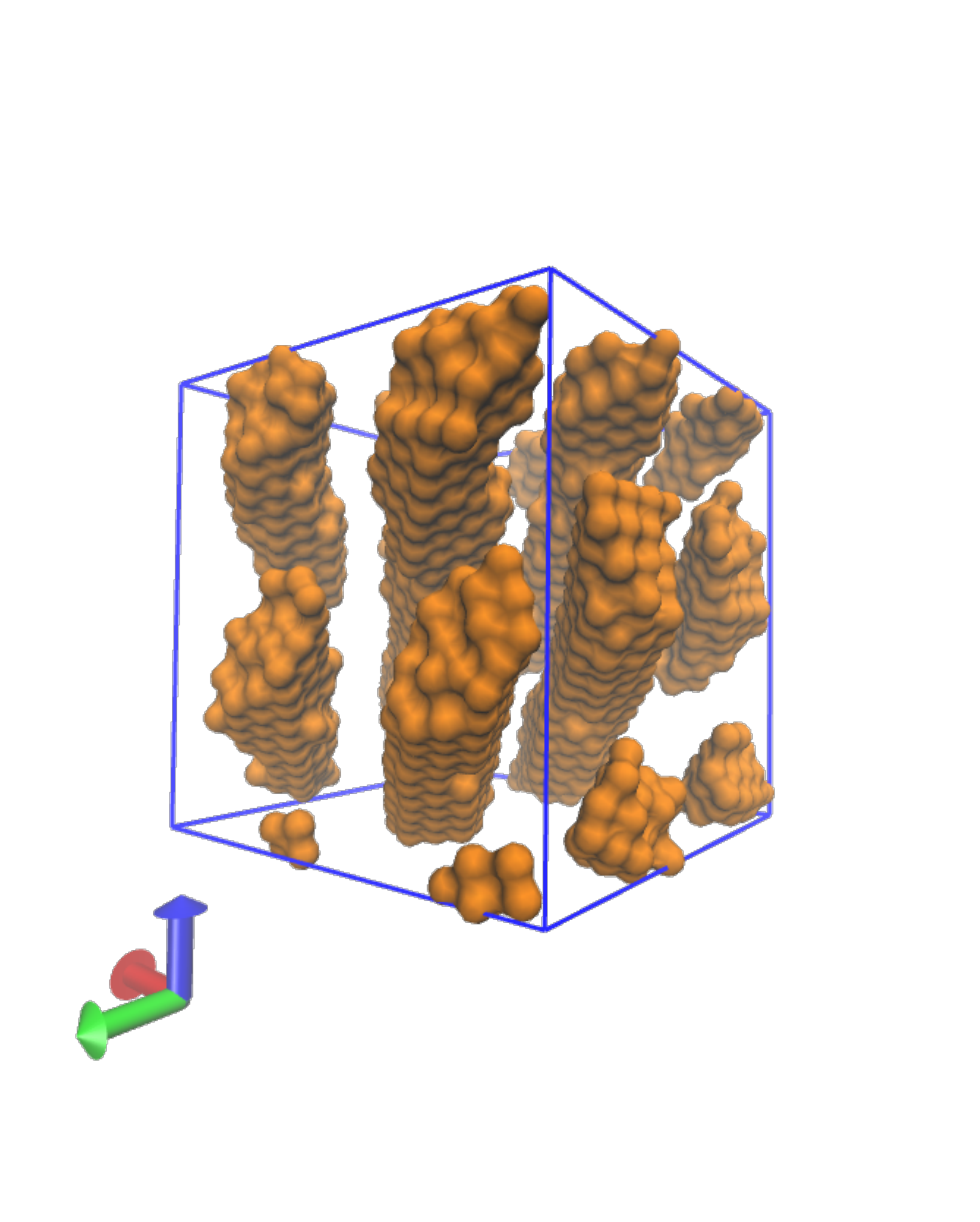}
}
\subfloat[$x=0.45$\label{fig:frame_rho04_x045}]{
\includegraphics[width=0.66\columnwidth]
{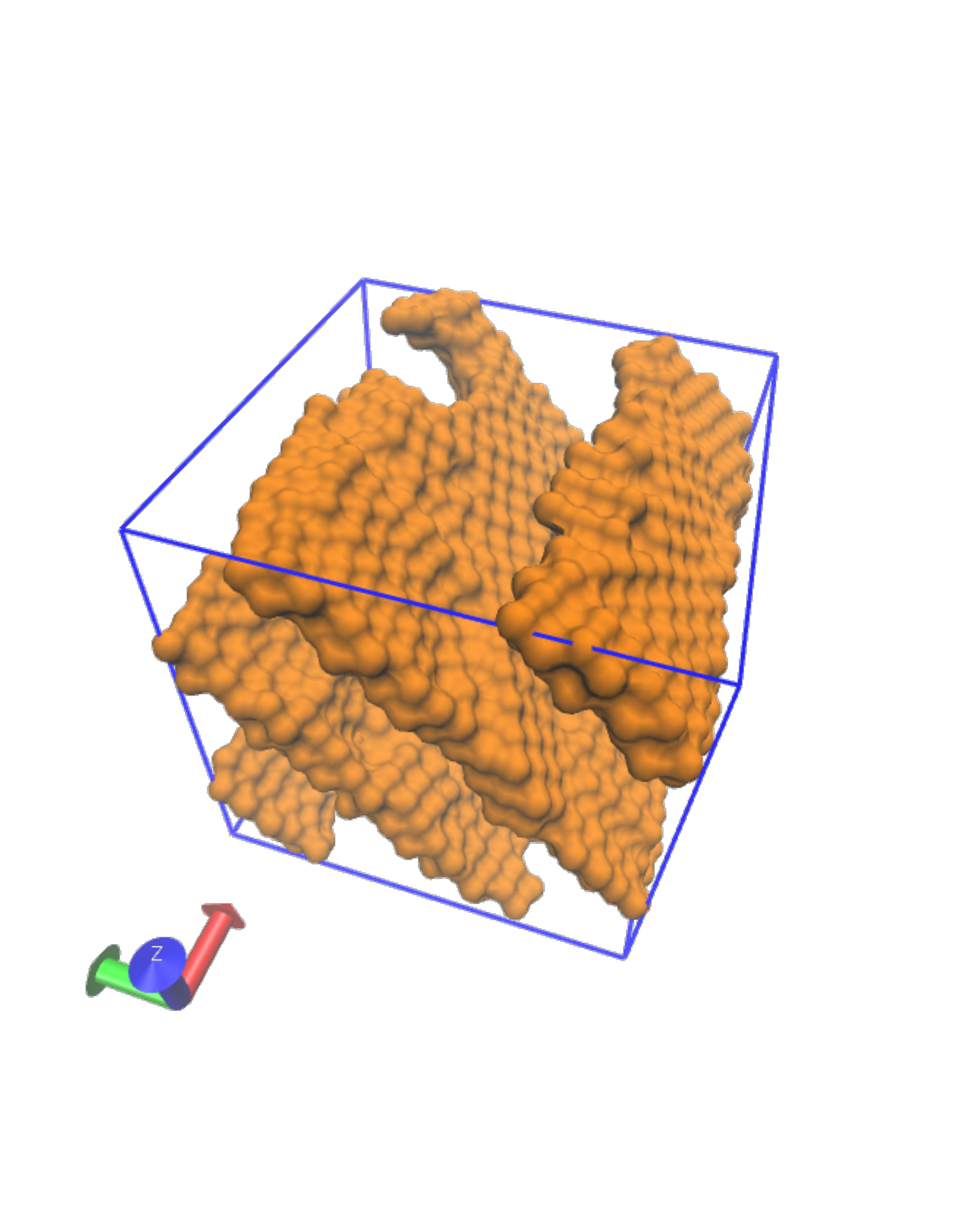}
}
\subfloat[$x=0.5$\label{fig:frame_rho04_x05_b}]{
\includegraphics[width=0.66\columnwidth]
{./lammps_rho04_x05_protons_surf_star.pdf}
}
\caption{\label{fig:frame_rho04} (color online)  Surface representation for the 
protons at the indicated fractions. The snapshot was taken at $T=0.2$ and 
$\rho=0.04$ for a system of $4000$ nucleons.  }
\end{figure*}

At lower densities, however, the situation is different. 
Figure~\ref{fig:frame_rho04} shows the protons for the cases of $x=0.1$, 0.2, 
0.3, 0.4, 0.45 and 0.5 at $T=0.2$ MeV and $\rho=0.04$ fm$^{-3}$. Comparing 
Figures~\ref{fig:frame_rho04_x02}, \ref{fig:frame_rho04_x04} and 
\ref{fig:frame_rho04_x05_b} to the the left end ($T=0.2$ MeV) values of $\chi$ 
obtained in Figure~\ref{fig:minkowski_asym} it is easy to see the how the voids 
and tunnels decrease as $x$ goes from 0.2 to 0.5.\\

\begin{figure*}[!htbp]
\centering
\subfloat[$\rho=0.02$\label{fig:gr_rho01_all}]{
\includegraphics[width=\columnwidth]
{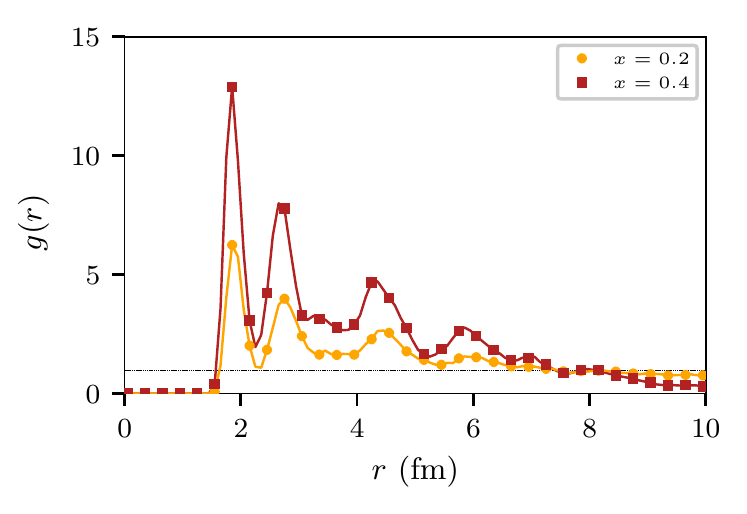}
}
\subfloat[$\rho=0.085$\label{fig:gr_rho085_all}]{
\includegraphics[width=\columnwidth]
{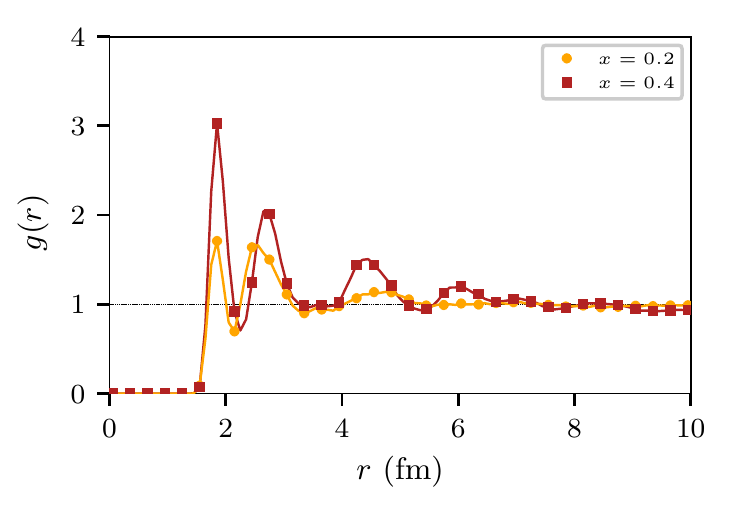}
}
\caption{\label{fig:gr_rho_asym} (color online) Radial distribution function 
$g(r)$ for nucleons corresponding to non-symmetric neutron star matter systems 
with $N=4000$ nucleons at $T=0.2\,$MeV. The system proton fraction is indicated 
in the inset. The
black horizontal line is a view guide for $g(r)=1$. All the curves correspond to
the New Medium model (see Section~\ref{NSP}). (a) Density $\rho=0.02$. (b)
Density $\rho=0.085$.  }
\end{figure*}

Further details can be understood by looking at the radial distribution 
function. Fig.~\ref{fig:gr_rho_asym} shows the $g(r)$ for two representative 
proton fractions at opposed densities. Notice from Fig.~\ref{fig:gr_rho085_all} 
that no qualitative differences can be
distinguished between the $x=0.2$ and the $x=0.4$ situations. Neither can be
distinguished between these and the symmetric case shown in 
Figure~\ref{fig:gr_rho02_rho085_all} (for $\rho=0.085$), except that the peaks 
appear somewhat sharper. Thus, the overall topology does not experience relevant 
changes for decreasing proton fractions at the highest explored density.\\

Figure~\ref{fig:frame_rho085_asym} shows three snapshots for the protons alone
at $\rho=0.085\,$fm$^{-3}$. The hollow spaces in there actually correspond to
regions occupied by neutrons (not represented for the sake of clarity). It can
be seen that the number of hollow regions (that is, the neutron regions) 
increases for decreasing fractions, while inside the ``proton regions'' no 
relevant changes can be noticed. This explains why the overall $g(r)$ does not 
exhibit qualitative changes in Fig.~\ref{fig:gr_rho085_all}, while 
Fig.~\ref{fig:eos_5} shows the ``right-branch'' pattern.\\

The radial distribution for the lower density $\rho=0.02\,$fm$^{-3}$ is 
qualitatively different from those at $\rho=0.085\,$fm$^{-3}$, as can be seen in 
Fig.~\ref{fig:gr_rho01_all}. The profiles appear sharply concentrated along the 
first few femtometers, while they gradually vanish towards $r\sim10\,$fm. Both 
profiles in Fig.~\ref{fig:gr_rho01_all} correspond to \textit{gnocchi}-like 
structures, similar to the one shown in Fig.~\ref{fig:frame_rho01_x05} for 
symmetric matter (not shown). Notice, however, that as the proton fraction 
diminishes, the profiles become somewhat smoother (see 
Fig.~\ref{fig:gr_rho01_all} and Fig.~\ref{fig:gr_rho02_rho085_all}).  \\

\begin{figure*}[!htbp]
\centering
\subfloat[$x=0.2$\label{fig:gr_rho01_x02_separated}]{
\includegraphics[width=\columnwidth]
{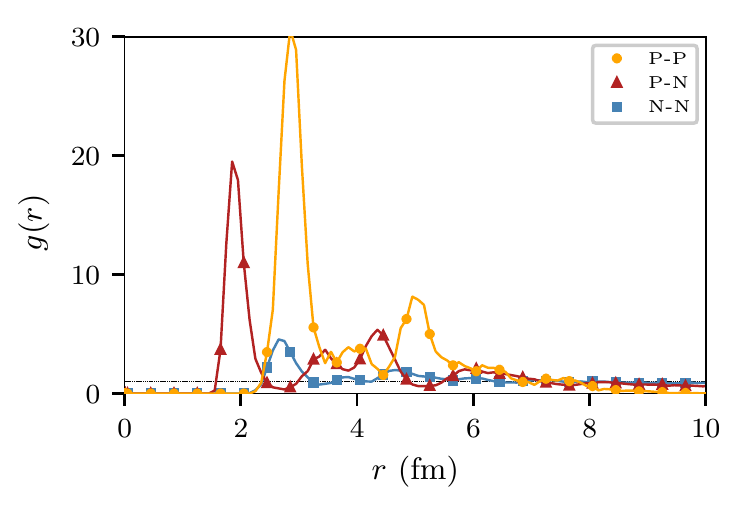}
}
\subfloat[$x=0.4$\label{fig:gr_rho01_x04_separated}]{
\includegraphics[width=\columnwidth]
{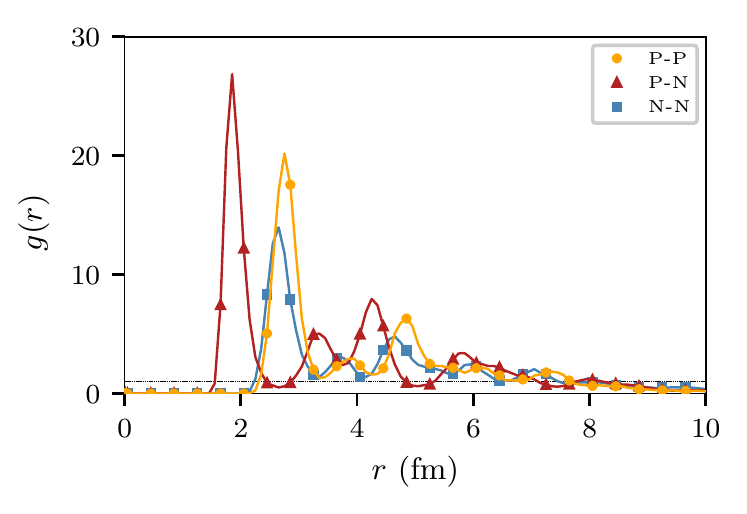}
}
\caption{\label{fig:gr_rho01_asym_separated} (color online) Radial distribution 
function $g(r)$ for nucleons corresponding to non-symmetric neutron star matter 
at $T=0.2\,$MeV, $\rho=0.01\,$fm$^{-3}$, and (a) $x=0.2$ and (b) $x=0.4$. The 
total number of nucleons is $N=4000$. In the inset ``P-P'' stands for 
proton-proton distances only, ``P-N'' corresponds to proton-neutron distances 
only, and ``N-N'' means neutron-neutron only.  }
\end{figure*}

Figure~\ref{fig:gr_rho01_asym_separated} splits Figure~\ref{fig:gr_rho01_all} 
into the corresponding profiles for protons and neutrons (see caption for 
details). It becomes clear from the comparison between 
Figures~\ref{fig:gr_rho01_x02_separated} and \ref{fig:gr_rho01_x04_separated} 
that the smoothening process in Figure~\ref{fig:gr_rho01_all} for decreasing 
proton fractions is related to the $g(r)$ distribution of the species, within 
the \textit{gnocchi} structure (say, $r\leq6\,$fm). Specifically, the 
proton-neutron (P-N) and neutron-neutron (N-N) distributions smoothen when the 
fraction $x$ diminishes. Notice, however, that the proton-proton distribution 
$g(r)$ sharpens around $3\,$fm. But this does not balance the P-N and N-N 
smoothening because of the reduced number of protons for $x=2$.\\

We may further examine the simulation cell for the intermediate density
$\rho=0.04\,$fm$^{-3}$. Fig.~\ref{fig:frame_rho04} exhibits the corresponding
proton structures (neutron not shown) for a sequence of fractions $x<0.5$. We
can see that as the proton fraction decreases, the ``pasta'' splits into
smaller pieces. The ``pasta'' structures for nearly symmetric neutron
star matter ($x\simeq0.5$) presents as \textit{lasagna}-like structures, while
the low proton fraction scenario ($x<0.3$) forms \textit{gnocchi}-like
structures. The \textit{spaghetti}-like structures appear in between
($x\simeq0.4$).  \\

From the comparison between Fig.~\ref{fig:eos_x} and Fig.~\ref{fig:frame_rho04}
we confirm that the changing topology of the system is due to the $\cup$-shape
pattern for the internal energy as a function of $x$. The 
\textit{spaghetti}-like structure achieves the minimum energy. The 
\textit{lasagnas} and the \textit{gnocchis} correspond
to higher energies on either branch of the $\cup$-shape.\\

In summary, two major effects appear when departing from symmetry. A
topological re-arrangement of the pasta structures occurs, attaining some
kind of fragmentation as the proton fraction diminishes. The lowest explored
proton fraction attains the \textit{gnocchi} structure, but, the
\textit{gnocchis} themselves experience (inner) topological changes during its
formation process. The resulting energy level surpasses the one at
$x\sim0.3$ in agreement with the observations made in reference to the energy 
$\cup$-shape dependence on $x$.

\subsection{\label{subsec:sym_energy_3}The Symmetry energy}

We pointed out in Sections~\ref{subsec:asym_energy} and
\ref{subsec:asym_topology} that topological changes taking place for
decreasing proton fractions are indeed related to the $\cup$-shape dependence of 
the energy on $x$ shown in Figure~\ref{fig:eos_x}. These patterns also carry 
information on the symmetry energy contribution to the equation of state; in 
this section we calculate $E_{syn}$ according to the procedure detailed in 
Section~\ref{subsec:sym_energy_3} (and Appendix~\ref{fittings}).\\

\begin{figure*}[!htbp]
\centering
\subfloat[$\rho=0.04$\label{fig:fitting_rho04}]{
\includegraphics[width=\columnwidth]
{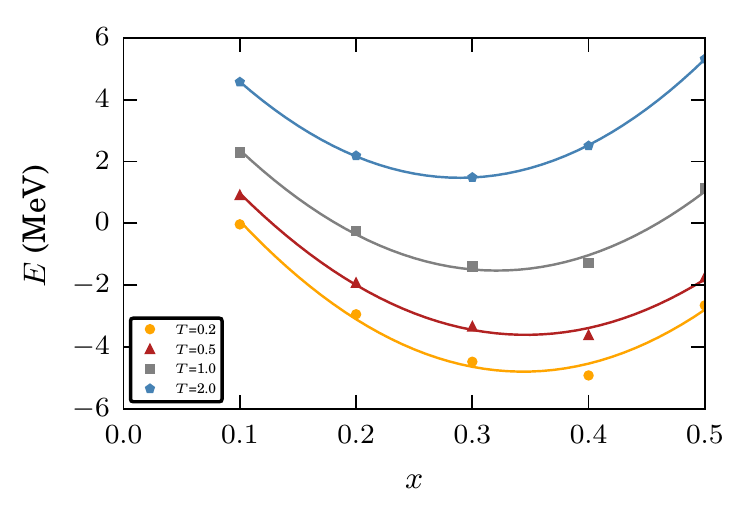}
}
\subfloat[$\rho=0.085$\label{fig:fitting_rho085}]{
\includegraphics[width=\columnwidth]
{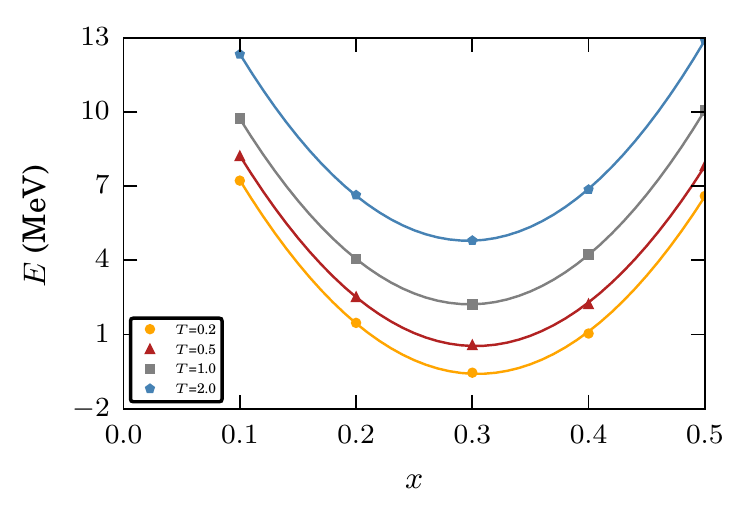}

}
 \caption{\label{fig:fitting_rhos} (color online) Internal energy per nucleon 
for neutron star matter at the indicated densities for systems with $N=4000$ 
nucleons interacting through the New Medium model. The rounded symbols 
correspond to CMD data, and the lines correspond to the estimated values after 
the two-step fitting procedure mentioned in Section~\ref{subsec:sym_energy_3} 
and Appendix~\ref{fittings}.}
\end{figure*}

Fig.~\ref{fig:fitting_rhos} shows the fittings for the internal energy as a
function of the proton fraction $x$. The curves represent the expected internal
energy, according to Equation~(\ref{fitting_2}), although the term of
$\mathcal{O}(\alpha^4)$ was neglected for simplicity (see
Appendix~\ref{fittings} for details). Data appears to fit quite well into this
quadratic profile. However, the low temperature curves shown in
Fig.~\ref{fig:fitting_rho04} exhibit some kind of departure from data at
$x\sim0.4$. This is somewhat an $\mathcal{O}(\alpha^4)$ discrepancy since a
noticeable improvement can be obtained if an $\alpha^4$ term is added
into the fitting procedure (not shown).\\

The fittings for the highest explored density (see
Fig.~\ref{fig:fitting_rho085}) match better the CMD data than the fittings at
$\rho=0.04$ (Fig.~\ref{fig:fitting_rho04}). Recall that the former correspond
to more compact proton structures (see Fig.~\ref{fig:frame_rho085_asym}) than
the latter (see Fig.~\ref{fig:frame_rho04}). Thus, as already noticed in
Section~\ref{subsec:asym_energy} and and \ref{subsec:asym_topology}, more
``fragmented'' structures (say, \textit{spaghettis} instead of
\textit{lasagnas}) flattens the right-branch of the U-shape pattern in
Fig.~\ref{fig:fitting_rho04}, departing from a seemingly quadratic profile. The
fittings discrepancy expresses this issue.\\

\begin{figure*}[!htbp]
\centering
\subfloat[nuclear matter\label{fig:esym_nsm_nm}]{
\includegraphics[width=\columnwidth]
{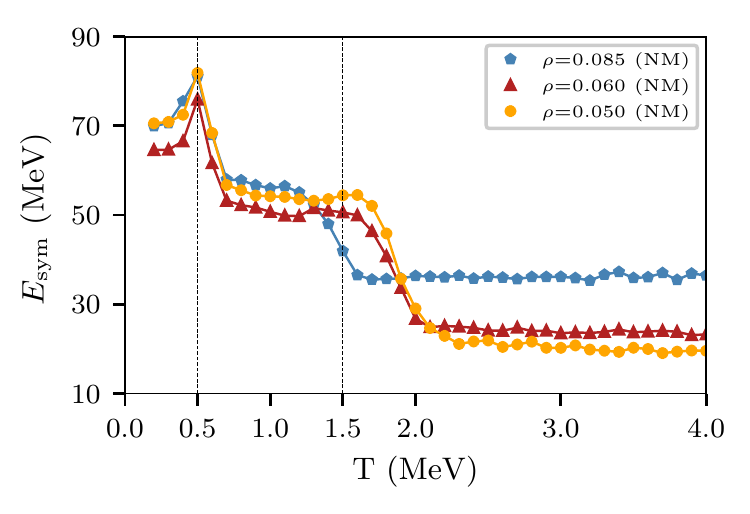}
}
\subfloat[neutron star matter\label{fig:esym_non_shifted}]{
\includegraphics[width=\columnwidth]
{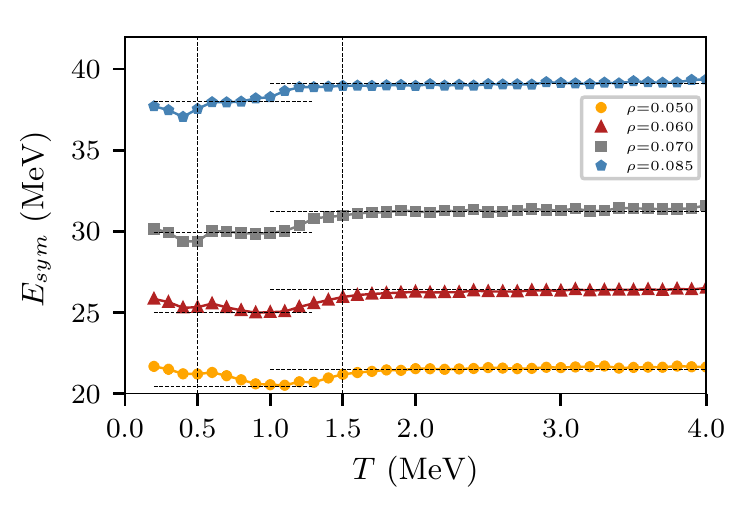}

}
 \caption{\label{fig:esym_all_densities} (color online) Symmetry energy as a
function of the temperature for the densities indicated in the insert in 
fm$^{-3}$. (a) Symmetry energy for nuclear matter (Pandharipande Medium 
Model) as a function of the temperature (extracted from 
Ref.~\cite{dorso2018_arxiv}). The vertical dashed lines are a guide to the eye. 
  (b) Symmetry energy for $N=4000$ nucleons embedded in an electron cloud, as 
explained in Section~\ref{NSP}. The fitting procedures are detailed in 
Section~\ref{subsec:sym_energy_3} and Appendix~\ref{fittings}. The horizontal 
and vertical dashed lines are a guide to the eye.   }
\end{figure*}

We further proceed to compute the $E_{sym}$ for neutron star
matter (within the explored range). Fig.~\ref{fig:esym_non_shifted} shows
the computed $E_{sym}$ for two different densities. The corresponding $E_{sym}$ 
for nuclear matter (Medium Model) appearing in Ref.~\cite{dorso2018_arxiv} has 
also been included for comparison. Notice that the $E_{sym}$ levels for nuclear 
matter and neutron star matter are essentially the same above $T\sim 2\,$MeV, 
but differ for lower temperatures. The upper regime (above $2\,$MeV) confirms 
that the $E_{sym}$ has been computed accurately, since either the internal 
energy in Fig.~\ref{fig:fitting_rho04} and the $E_{sym}$ match the expected 
values.\\

It can be seen in Fig.~\ref{fig:esym_non_shifted} a smooth change in
the profile slopes for temperatures below $1.5\,$MeV. This is (roughly) in
agreement with the patterns shown in Fig.~\ref{fig:esym_nsm_nm} for the nuclear
matter (Medium) model, although the slopes' signs are different. The neutron 
star matter profiles appear (qualitatively) inverted with respect to the 
nuclear matter profiles (say, the slopes present opposite signs). \\

The estimation with the $\mathcal{O}(\alpha^4)$ term fits better the
CMD data exhibited in Fig.~\ref{fig:eos_rho} for $x\sim 0.4$ and for the lowest 
explored temperature (not shown). However, the $E_{sym}$ level at higher 
temperatures (above $1.5\,$MeV) spreads out, not allowing a precise estimation 
of the corresponding energy level. Thus, the quadratic order estimations shown 
in Fig.~\ref{fig:esym_all_densities} are somehow more practical. \\

The major result from the above observations is that the $E_{sym}$ experiences
a change in its slope at the pasta regime, driving the energy level to lower
values for decreasing temperatures (below $1.5\,$MeV). This is a somewhat
opposite behavior with respect to the known results for the nuclear matter model
(see Ref~\cite{dorso2018_arxiv}). Furthermore, this behavior holds for either
fittings of order $\alpha^2$ and $\alpha^4$.  \\

A second result can be outlined from the widely explored range of the proton
fraction. The ``left-branches'' appearing in Fig.~\ref{fig:fitting_rhos} handle
somehow the $E_{sym}$ level at the pasta regime (below $2\,$MeV). But, as
mentioned in Section~\ref{subsec:asym_topology}, this ``left-sided'' behavior
accomplishes topological changes within the (fragmented) pasta structures at
low densities (within the explored range). Thus, the re-arrangements in the
(fragmented) pasta structures (say, \textit{gnocchi}-like or
\textit{spaghetti}-like) themselves are able to switch the $E_{sym}$ level up
or down with respect to the level above $2\,$MeV. In brief, the 
presence of pasta influences the behavior of the $E_{sym}$ both for NM and 
NSM.\\

\section{Conclusions}\label{concluding}

In this article we have studied the behavior of Neutron Star matter (according 
to the CMD approach). At rather low temperatures ($0.2\,$MeV$\leq 
T\leq$4$\,$MeV) and densities in the range of 0.02$\,$fm$^{-1}$ to 
0.085$\,$fm$^{-1}$ for isospin symmetric ($x=0.5$) and non-symmetric 
($x=0.1-0.4$). We have investigated the formation of ``pastas'', the emergence 
of phase transitions and the behavior of the symmetry energy ($E_{sym}$). Using 
different techniques of molecular dynamics we studied ``infinite systems'' 
using primordial cell of 4000 nucleons and periodic boundary conditions. \\

We calculated the caloric curve and found that as temperature was lowered from 
4$\,$MeV at a given point (irrespective of the value of $x$ in the studied 
range) a clear change of the slope takes place, signaling the break down of the 
homogeneous situation as the system undergoes the topological phase transition. 
If the temperature is further lowered the signal of the first order (liquid to 
solid) phase transition is clearly seen. At variance with the case of the 
nuclear matter, the caloric curves (CC) for the different values of $\rho$ 
remain parallel. \\

The behavior of the system was also analyzed using the Euler functional 
$\chi$. For temperatures above $T\simeq 1\,$MeV, $\chi$ increases with density. 
If the densities are $\rho=0.05-0.06\,$fm$^{-1}$, $\chi$ goes from negative 
values (tunnel dominated configurations) to positive ones (configurations 
dominated by cavities). \\

All these phase changes that were detected as the temperature was lowered are 
reflected in the behavior of the symmetry energy. It is clearly seen that for 
temperatures above 2$\,$MeV, the symmetry energy calculated according to the 
method explained in the appendix remains basically constant. On the other hand, 
 for temperatures around the values for which we have changes in the slope of 
the caloric curves the symmetry energy also displays changes, as was stated for 
the case of nuclear matter.\\

In conclusion, classical molecular dynamics simulations show the formation of 
``pastas'' in the isospin symmetric and non-symmetric nuclear star matter. The 
behavior of $E_{sym}$, together with the two phase transitions are an 
indication of interesting phenomena that are present in the neutron star crust 
and supernova explosions.

\begin{acknowledgments}
Part of this study was financed by FONCyT (Fondo para la
Investigaci\'on Cient\'\i fica y Tecnol\'ogica) and Inter-American Development
Bank (IDB),  Grant Number PICT 1692 (2013).
\end{acknowledgments}

\appendix

\section{\label{fittings}The symmetry energy fitting procedure}

The fitting procedure outlined in Section~\ref{esymm} corresponds to a two
step-fitting method. The coefficients $E_i$ ($i=0...3$) from
Eq.~(\ref{fitting_1}) are computed at the first stage of the procedure. These
coefficients, however, depend on the fraction $\alpha$, although not directly
on the fraction $x$ (recall that $\alpha=1-2x$). A more suitable (low
order) expression for the $E_i$'s instead of Eq.~(\ref{fitting_2}) is

\begin{equation}
\begin{array}{lcl}
         E_i(T,x) & = &
E_{i0}+E_{i2}\,\overbrace{(1-4x+4x^2)}^{\alpha^2}+\mathcal{O}(\alpha^4)\\
        & & \\
        & \simeq &
(E_{i0}+E_{i2})-4E_{i2}\,x+4E_{i2}\,x^2\\
        \end{array}\label{fitting_3}
\end{equation}

The above expression corresponds to the lowest non-trivial expansion
$E_i(T,x)\simeq \tilde{E}_{i0}+\tilde{E}_{i1}\,x+\tilde{E}_{i2}\,x^2$. Thus,
the $\tilde{E}_{ij}$'s ($j=0,1,2$) are related to the $E_{i0}$, $E_{i2}$
coefficients according to the following matrix relation\\

\begin{equation}
 \left(\begin{array}{cc}
        1  & 1 \\
            &  \\
         0  & -4  \\
          &   \\
         0  & 4  \\
        \end{array}\right)\,\left(\begin{array}{c}
                                  E_{i0} \\
                                  \\
                                      E_{i2} \\
                                   \end{array}\right)=\left(\begin{array}{c}
                                     \tilde{E}_{i0}\\
                                     \\
                                     \tilde{E}_{i1} \\
                                     \\
                                     \tilde{E}_{i2} \\
                                   \end{array}\right)\label{fitting_4}
\end{equation}

Notice that $\tilde{E}_{i0}$, $\tilde{E}_{i1}$, $\tilde{E}_{i2}$
actually correspond to the fitting parameters at the second stage of the
procedure outlined in Section~\ref{esymm} (that is, after the $E_i$'s were
obtained). The  $E_{i0}$ and $E_{i2}$ parameters, though, are now computed
indirectly from the overdetermined system (\ref{fitting_4}). The best
estimates for $E_{i0}$ and $E_{i2}$ that we can think about correspond to the
least squares estimation. Thus, the estimate for $E_{i2}$ gives \\

\begin{equation}
E_{i2}=\displaystyle\frac{\tilde{E}_{i2}-\tilde{E}_{i1}}{8}
\pm\displaystyle\frac{1}{\sqrt{2}}\,\displaystyle\frac
{\tilde{E}_{i2}-\tilde{E}_{i1}}{8}\label{fitting_5}
\end{equation}

\noindent where the first term on the right corresponds to the estimated
(least squares) mean value, while the second term represents the standard
deviation. \\

The resulting estimate for $E_{i2}$ in (\ref{fitting_5}) is actually the
mid-value between the (incompatible) solutions $-\tilde{E}_{i1}/4$ and
$\tilde{E}_{i2}/4$ that can be obtained (independently) from the second and
third rows in (\ref{fitting_4}), respectively. Thus, the estimation
(\ref{fitting_5}) may be envisaged as the resultant for equally weighting the
linear and the quadratic terms in Eq.~(\ref{fitting_3}). This is the key
consequence of introducing the substitution $\alpha\rightarrow x$ in
Eq.~(\ref{fitting_3}). \\

Our numerical computations for $E_{i2}$ at the examined temperatures and
densities indicated in Section~\ref{subsec:sym_energy_3} are detailed in
Table~\ref{table_esym}.\\

\begin{table}[!htbp]
{
\begin{tabular}{c@{\hspace{17mm}}r@{\hspace{9mm}}r@{\hspace{9mm}}r@{\hspace{9mm
}}r}
\toprule
$T$ & $E_{02}$ & $E_{12}$ & $E_{22}$ & $E_{32}$ \\
\colrule
0.2   & -9.472    &  976  & -10080 & 60305\\
0.5   &  -35.721  &  2249 & -30633 & 168400\\
1.0   &   0.745   &  408  & -1137 & 18048\\
2.0   &   3.401   &  239  &  2985 & -10323\\
\botrule
\end{tabular}
}
\caption{Least squares estimates according to Eq.~(\ref{fitting_5}). The
standard deviations may be obtained easily from these
estimates and Eq.(~\ref{fitting_5}). }
\label{table_esym}
\end{table}

\end{document}